\documentclass[aps,prc,reprint,superscriptaddress, amsmath,amssymb,nofootinbib]{revtex4-2}
\usepackage{graphicx}
\usepackage{dcolumn}
\usepackage{bm}
\usepackage{hyperref}
\usepackage{tabularx}
\usepackage{longtable} 
\usepackage{amsmath} 

\begin{document}

\title{Shapes, Softness and Non-Yrast Collectivity in $^{\textbf{186}}$W}


\author{V.~S.~Prasher}
\affiliation{Department of Physics and Applied Physics, University of Massachusetts Lowell, Lowell, Massachusetts 01854, USA}

\author{A.~J.~Mitchell}
\email[]{aj.mitchell@anu.edu.au}
\affiliation{Department of Physics and Applied Physics, University of Massachusetts Lowell, Lowell, Massachusetts 01854, USA}
\affiliation{Department of Nuclear Physics, Research School of Physics, The Australian National University, Canberra, ACT 2601, Australia}

\author{C.~J.~Lister}
\affiliation{Department of Physics and Applied Physics, University of Massachusetts Lowell, Lowell, Massachusetts 01854, USA}

\author{P.~Chowdhury}
\affiliation{Department of Physics and Applied Physics, University of Massachusetts Lowell, Lowell, Massachusetts 01854, USA}

\author{L.~Afanasieva}
\affiliation{Department of Physics and Astronomy, Louisiana State University, Baton Rouge, Louisiana 70803, USA}

\author{M.~Albers}
\affiliation{Physics Division, Argonne National Laboratory, Argonne, Illinois 60439, USA}

\author{C.~J.~Chiara}
\altaffiliation[Present address: ]{DEVCOM Army Research Laboratory, Adelphi, Maryland 20783, USA}
\affiliation{Physics Division, Argonne National Laboratory, Argonne, Illinois 60439, USA}
\affiliation{Department of Chemistry and Biochemistry, University of Maryland, College Park, Maryland 20742, USA
}

\author{M.~P.~Carpenter}
\affiliation{Physics Division, Argonne National Laboratory, Argonne, Illinois 60439, USA}

\author{D.~Cline}
\affiliation{Nuclear Structure Research Laboratory, University of Rochester, Rochester, New York 14627, USA}

\author{N.~D'Olympia}
\altaffiliation[Present address: ]{The MITRE Corporation, Bedford, Massachusetts 01730, USA}
\affiliation{Department of Physics and Applied Physics, University of Massachusetts Lowell, Lowell, Massachusetts 01854, USA}

\author{C.~J.~Guess}
\altaffiliation[Present address: ]{Department of Physics and Astronomy, Rowan University, Glassboro, New Jersey 08028, USA}
\affiliation{Department of Physics and Applied Physics, University of Massachusetts Lowell, Lowell, Massachusetts 01854, USA}

\author{A.~B.~Hayes}
\affiliation{National Nuclear Data Center, Brookhaven National Laboratory, Upton, New York 11973, USA}

\author{C.~R.~Hoffman}
\affiliation{Physics Division, Argonne National Laboratory, Argonne, Illinois 60439, USA}

\author{R.~V.~F.~Janssens}
\affiliation{Department of Physics and Astronomy, University of North Carolina at Chapel Hill, Chapel Hill, North Carolina 27599-3255, USA}
\affiliation{Triangle Universities Nuclear Laboratory, Duke University, Durham, North Carolina 27708-2308, USA}

\author{B.~P.~Kay}
\affiliation{Physics Division, Argonne National Laboratory, Argonne, Illinois 60439, USA}

\author{T.~L.~Khoo}
\affiliation{Physics Division, Argonne National Laboratory, Argonne, Illinois 60439, USA}

\author{A.~Korichi}
\affiliation{Physics Division, Argonne National Laboratory, Argonne, Illinois 60439, USA}
\affiliation{IJCLab-IN2P3, F-91405 Orsay Campus, France}

\author{T.~Lauritsen}
\affiliation{Physics Division, Argonne National Laboratory, Argonne, Illinois 60439, USA}

\author{E.~Merchan}
\altaffiliation[Present address: ]{Mass General Brigham, Boston, Massachusetts 02145, USA}
\affiliation{Department of Physics and Applied Physics, University of Massachusetts Lowell, Lowell, Massachusetts 01854, USA}

\author{Y.~Qiu}
\altaffiliation[Present address: ]{Department of Physics and Astronomy, Shanghai Jiao Tong University, Shanghai 200240, China}
\affiliation{Department of Physics and Applied Physics, University of Massachusetts Lowell, Lowell, Massachusetts 01854, USA}

\author{D.~Seweryniak}
\affiliation{Physics Division, Argonne National Laboratory, Argonne, Illinois 60439, USA}

\author{R.~Shearman}
\affiliation{Department of Physics and Applied Physics, University of Massachusetts Lowell, Lowell, Massachusetts 01854, USA}
\affiliation{National Physical Laboratory, Teddington, Middlesex TW11 0LW, United Kingdom}
\affiliation{Department of Physics, University of Surrey, Guildford GU2 7XH, United Kingdom}

\author{S.~K.~Tandel}
\altaffiliation[Present address: ]{School of Physical Sciences, UM-DAE Centre for Excellence in Basic Sciences, University of Mumbai, Mumbai 400098, India}
\affiliation{Department of Physics and Applied Physics, University of Massachusetts Lowell, Lowell, Massachusetts 01854, USA}

\author{A.~Verras}
\altaffiliation[Present address: ]{Intuitive Machines, Houston, Texas 77058, USA}
\affiliation{Department of Physics and Applied Physics, University of Massachusetts Lowell, Lowell, Massachusetts 01854, USA}

\author{C.~Y.~Wu}
\affiliation{Lawrence Livermore National Laboratory, Livermore, California 94550, USA}

\author{S.~Zhu}
\affiliation{Physics Division, Argonne National Laboratory, Argonne, Illinois 60439, USA}
\affiliation{National Nuclear Data Center, Brookhaven National Laboratory, Upton, New York 11973, USA}


\date{\today}

\begin{abstract}

Non-yrast, excited states in neutron-rich $^{186}$W were populated via inelastic-scattering reactions using beams of $^{136}$Xe nuclei accelerated to 725 and 800 MeV. Levels populated in the reactions were investigated via particle-$ \gamma $ coincidence techniques using the Gammasphere array of High-Purity Germanium detectors and the compact heavy-ion counter, CHICO2. The $ K^{\pi} = 2 ^{+} $ ($\gamma$), $ K^{\pi} = 0^{+}$ and $ K^{\pi} = 2^{-} $ (octupole) rotational side bands were extended to spins $ 14\hbar $, $ 12\hbar $, and $ 13\hbar $, respectively. A staggering pattern observed in the energies of levels in the $ K^{\pi} = 2^{+} $ band was found to be consistent with a potential that gets softer to vibration in the $ \gamma $ degree of freedom with increasing spin. The odd-even staggering of states in the $ K^{\pi} = 2^{-}$ band was found to exhibit a phase opposite to that seen in the $ \gamma $ band; an effect most probably associated with Coriolis coupling to other, unobserved octupole vibrational bands in $^{186}$W.

\end{abstract}

\maketitle


\section{\label{introduction}INTRODUCTION}

The trajectory of nuclear shapes in rare-earth and transition elements between axially symmetric, prolate-deformed $_{~66}^{170}$Dy$_{104}$ \cite{regan2002} located at mid-shell and spherical, doubly magic $_{~82}^{208}$Pb$_{126}$ \cite{heusler2016} has long been predicted to pass through a region of nuclei with soft, triaxial shapes that evolve into oblate deformation as the proton, $ Z $, and (or)  neutron, $ N $, numbers increase, before reaching sphericity \cite{kumar1968}. The nuclear level structure and electromagnetic properties of Yb ($Z = 70$), Hf ($Z = 72$), W ($Z = 74$), Os ($Z = 76$) and Pt ($Z = 78$) isotopes have been subject to extensive experimental and theoretical study near stability, and exhibit these characteristics \cite{sariguren2008,stevenson2005}. However, information on the progression of shapes across the region is incomplete, especially for non-yrast modes of collective motion. 

The shape and softness of deformed nuclei can be revealed through detailed spectroscopy of $\gamma$-ray cascades induced by rotation of the mean field. While level spacings and lifetimes of ground-state-band members reveal the overall shape and collectivity, rotational side bands contain more nuanced information on softness to vibrations and axial asymmetry. This is most apparent in the even-even nuclei in this region, where pairing correlations act to lower the ground-state energies and push all non-collective, particle-hole states to about 2~MeV in excitation. In contrast, at low energies, the neighboring odd-$A$ nuclei are rich in complementary information on the Nilsson-like motion of unpaired valence particles \cite{casten1973}. 

\begin{figure}[t!]
\begin{center}
\includegraphics*[width=8.5cm]{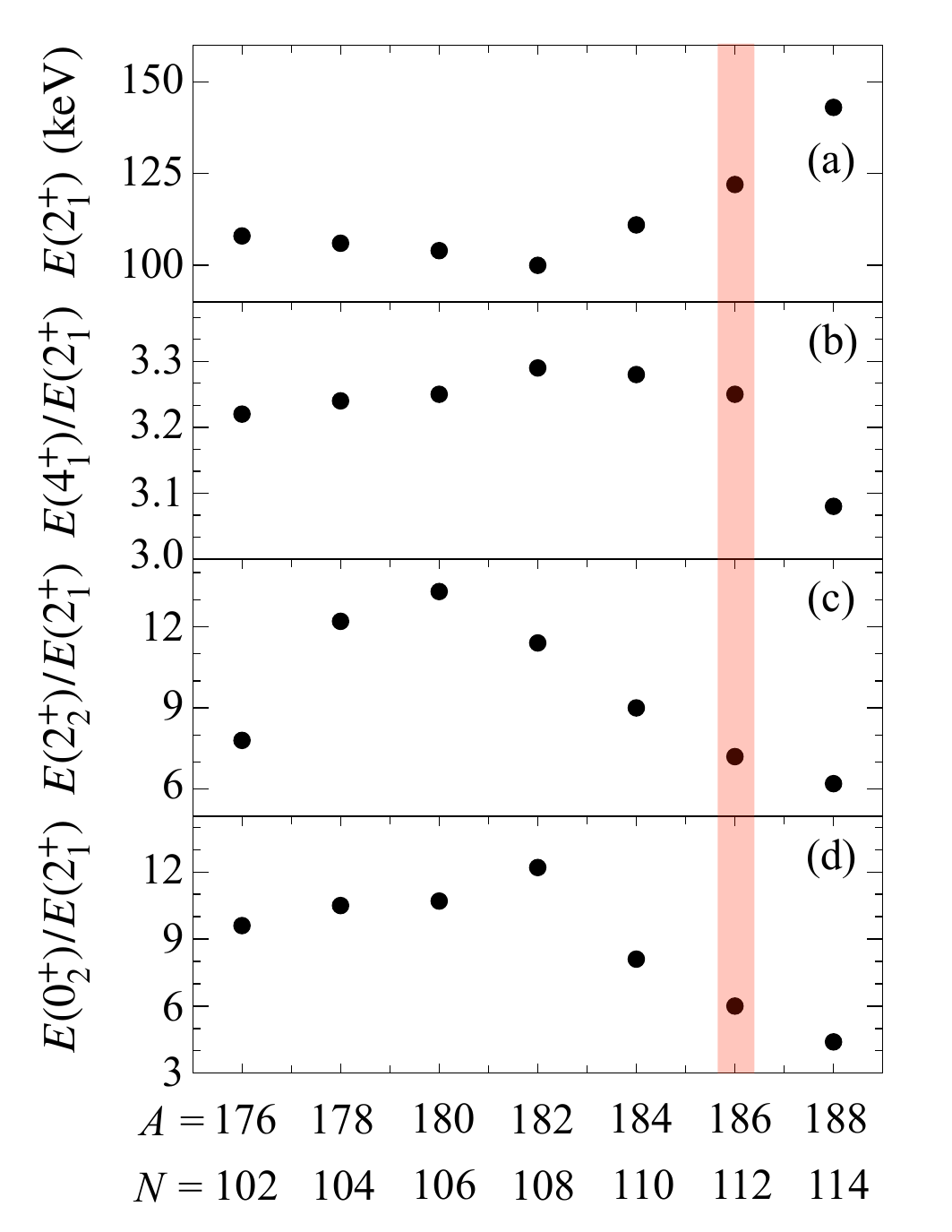} 
\caption{Systematic trends in collective parameters for even-$A$ tungsten nuclei, including (a) the first excited-state energy ($2^+_1$) and the ratios of the (b) $ 4^+_1$ level, (c) $K^{\pi} = 2^+$, $\gamma$-bandhead ($2^+_2$), and (d) $K^{\pi} = 0^+$ bandhead ($0^+_2$) energies to the first excited-state energy, E($2^+_1$) \cite{A=176,A=178,A=180,A=182,A=184,A=186,A=188}.}
\label{fig1}
\end{center}
\end{figure}

Nuclear-structure properties along the W isotopic chain have been studied for many years, using a variety of spectroscopic approaches (for example, see Refs.~\cite{A=176,A=178,A=180,A=182,A=184,A=186,A=188} and references therein). In the even-even isotopes, rotational bands have been established up to moderate spins and shape transitions have been identified, as well as $K$-isomeric states associated with well-deformed, axially symmetric, prolate shapes \cite{dracoulis2016, purry1998, wheldon1998, lane2010}. Figure~\ref{fig1} presents some key systematic trends in the tungsten isotopes between $ 176 \leq~A~\leq 188 $ \cite{A=176,A=178,A=180,A=182,A=184,A=186,A=188}. Energy systematics \cite{alkhomashi2009}, as well as known $B(E2; 2^+_1 \to 0^+_1)$ values \cite{rudigier2010, mason2013}, indicate that $^{186}$W, at $N$~=~112, lies beyond the maximum axial deformation and is softening in shape \cite{sariguren2008}.

Tungsten-186 is the heaviest stable W isotope. Its low-lying structure has been investigated using Coulomb excitation with proton, $^{4}$He, $^{16}$O, and $^{208}$Pb beams~\cite{mcgowan1977, milner1971, kulessa1989}; it was also studied following $\beta$-decay of the parent Ta isotope~\cite{monnand1969}. Experimental data on the non-yrast, higher-spin states are sparse due to the lack of any suitable heavy-ion fusion-evaporation reaction. 

The approach adopted in this study to reach the elusive higher-angular-momentum states of interest was the use of heavy-ion inelastic scattering. This method was successful in accessing non-yrast bands of neutron-rich $^{180}$Hf to high spin in an earlier study \cite{ngijoi-yogo2007,tandel2008-1}, where states of spin up to 20$\hbar$ were populated. In this work, we focus on spectroscopy of side bands in $^{186}$W. These have been extended to relatively high spin and enable an investigation of the shape softness and evolution with angular momentum. 

\section{\label{experiment} EXPERIMENTAL DETAILS}

Excited states in $^{186}$W \cite{A=186} were populated via ``inelastic'' scattering{\footnote[1]{At energies 10 and 20\% above the Coulomb barrier, reaction mechanisms are complex (see, for example Refs.~\cite{broda2006, broda2012} for a review). These are all regrouped here under the term ``inelastic" for simplicity.}} of $^{136}$Xe beams of 725 and 800~MeV (10 and 20$\%$ above the Coulomb barrier, respectively) delivered by the ATLAS accelerator at Argonne National Laboratory. The beams impinged upon a thin target of $^{186}$W ($99.8\%$ enriched) that was 250-$ \mu \rm{g/cm}^{2}$ thick and backed by a 110-$\mu\rm{g/cm}^{2}$ thick, carbon foil~\cite{prasher2015}. Scattered beam- and target-like ions were detected and identified with the upgraded Rochester-Livermore $4\pi$ compact heavy-ion counter, CHICO2~\cite{wu2016}. 

Prompt $\gamma$ rays emitted from excited states in the reaction partners were detected by the Gammasphere array, which was comprised of 91 Compton-suppressed, High-Purity Germanium (HPGe) detectors. The relative $\gamma$-ray detection efficiency of Gammasphere was determined using standard calibration sources of $^{152}$Eu and $^{243}$Am mounted in the CHICO2 target holder before and after the ``in-beam" measurements were performed.   

\begin{figure}[t!]
\begin{center}
   \includegraphics*[trim=0cm 5.7cm 0cm 0cm, clip=true, width=8.2cm]{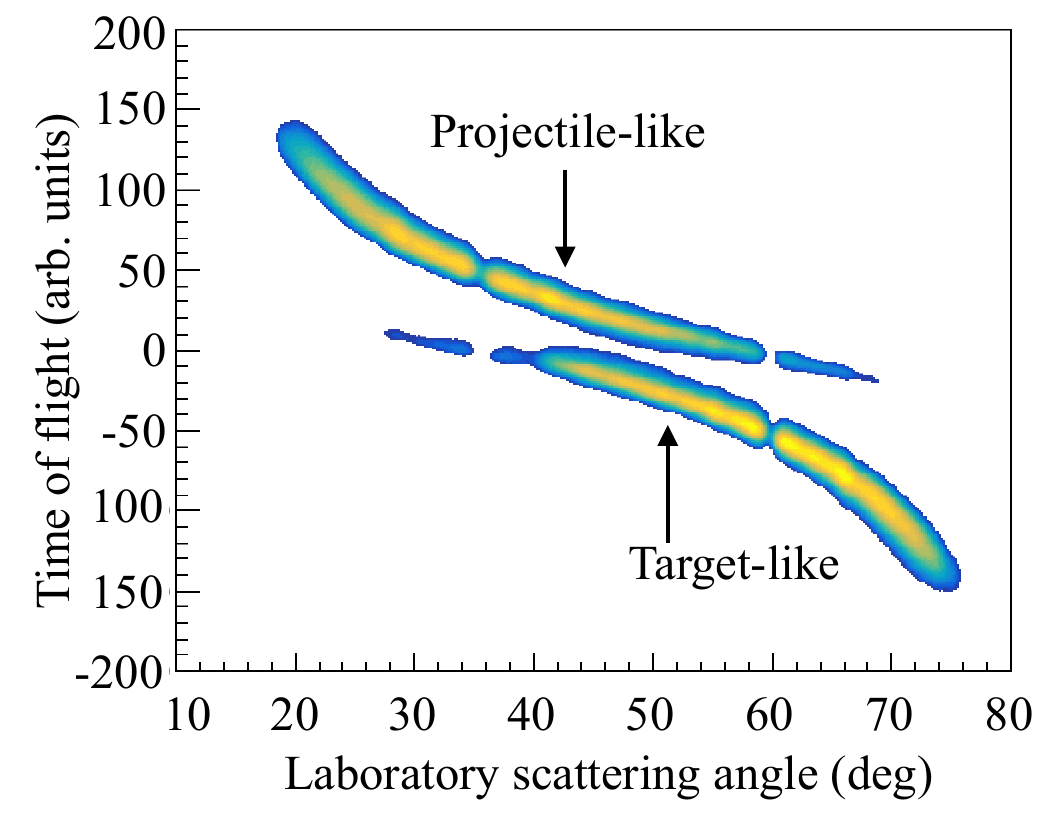} 
\caption{Histogram of the time-of-flight difference between projectile- and target-like fragments versus scattering angle ($\theta$), with a lower limit of 10 000 counts per channel displayed for clarity. The gaps in data at $\theta = 37^{\circ}$ and $59^{\circ}$ are due to support ribs for the CHICO2 pressure window~\cite{wu2016}.}
\label{fig2}
\end{center}
\end{figure}

\begin{figure}[t!]
\begin{center}
\includegraphics*[width=8.2cm]{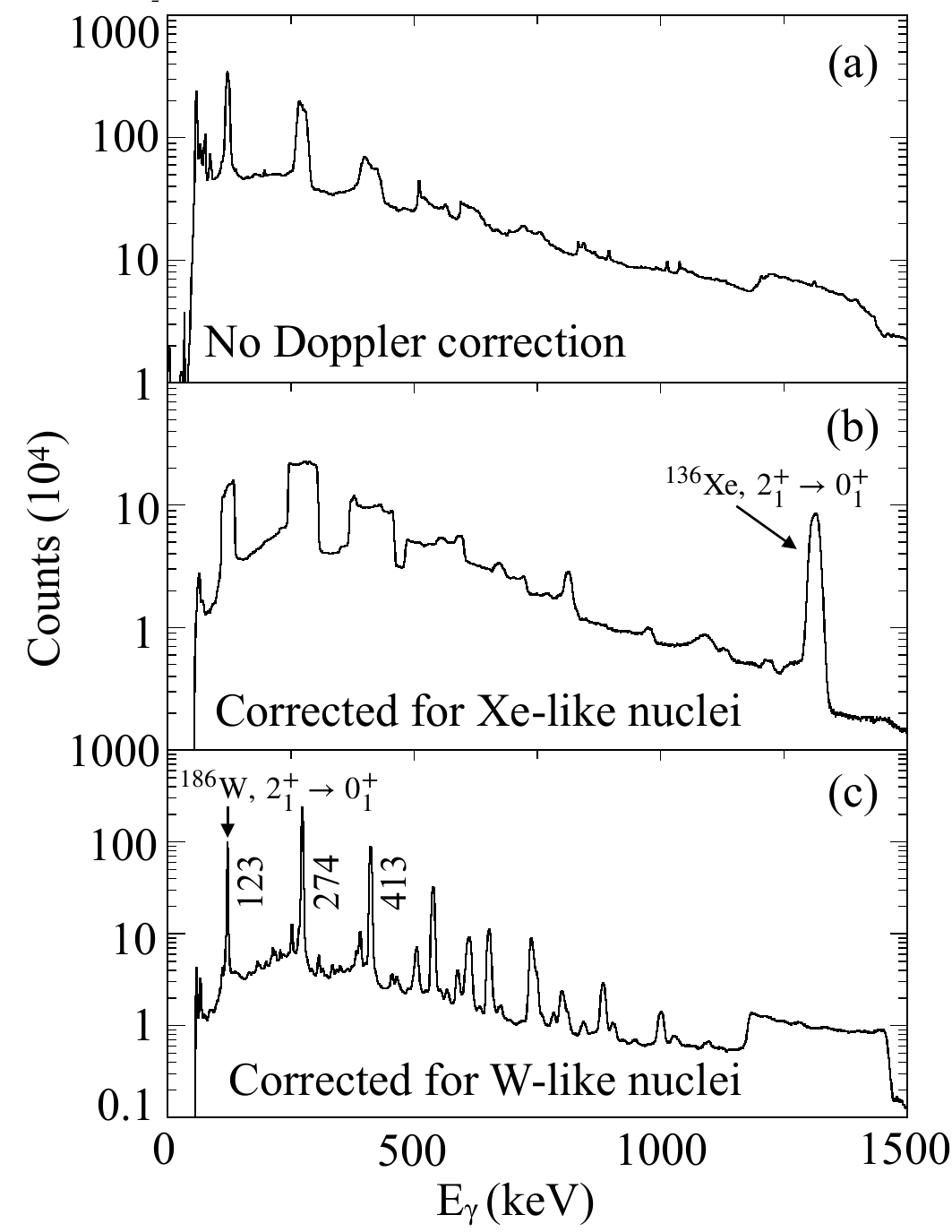} 
\caption{Total projection of the measured $\gamma$-ray energies (a) uncorrected for Doppler shifts, (b) corrected for beam-like nuclei and (c) corrected for target-like nuclei. Peaks corresponding to transitions in $^{186}$W are sharpened when the appropriate correction is applied for the target-like ion recoil velocity, broadened if the correction for the beam-like reaction partner is applied instead, and vice versa for $\gamma$ rays associated with beam-like nuclei such as the $ 2^+ \to 0^+ $ transition in $^{136}$Xe \cite{A=136} labeled in panel (b).}
\label{fig3}
\end{center}
\end{figure}

The data-acquisition system had a master trigger requirement of at least one prompt $\gamma$ ray being detected by Gammasphere in coincidence with two co-planar fragments measured in CHICO2. The particle position and time-of-flight difference determined from the CHICO2 information \cite{wu2016} were used to distinguish between beam- and target-like fragments following inelastic excitation or nucleon transfer (Fig.~\ref{fig2}). The deduced particle velocity and emission angle were then used to reconstruct particle kinematics event-by-event, which enabled energies of prompt $\gamma$ rays emitted by nuclei decaying in-flight to be corrected for Doppler shifts (Fig.~\ref{fig3}). In this work, the energies of known transitions and excited states \cite{A=186} were used to provide an internal calibration of the Doppler-corrected $\gamma$-ray energies. Uncertainties in energies of the new $\gamma$ rays identified here are approximately $\pm 0.3$~keV. 

Beam currents of $\approx 0.25$~pnA used in the experiments resulted in a trigger rate of $\approx 1.5$~kHz per CHICO2 element, and $\approx 1$~kHz per HPGe detector in Gammasphere. In total, approximately $3 \times 10^{8}$ CHICO2-triggered, Compton-suppressed events with two or more $\gamma$ rays were collected. The raw data events were stored on disk and sorted offline with the \textsc{DGSSort} program \cite{dgssort} in combination with the \textsc{Root} object-oriented framework \cite{brun1997}. The \textsc{Radware}~\cite{radware} programs $\textsc{Escl8r}$ and $\textsc{Levit8r}$ were used to inspect $\gamma-\gamma$ and $\gamma-\gamma-\gamma$ coincidence relationships, respectively, and to build the level scheme.

\section{\label{results}RESULTS}

\begin{figure*}[t]
\begin{center}
   \includegraphics*[trim=0cm 8.5cm 0cm 0cm, clip=true, width=17cm]{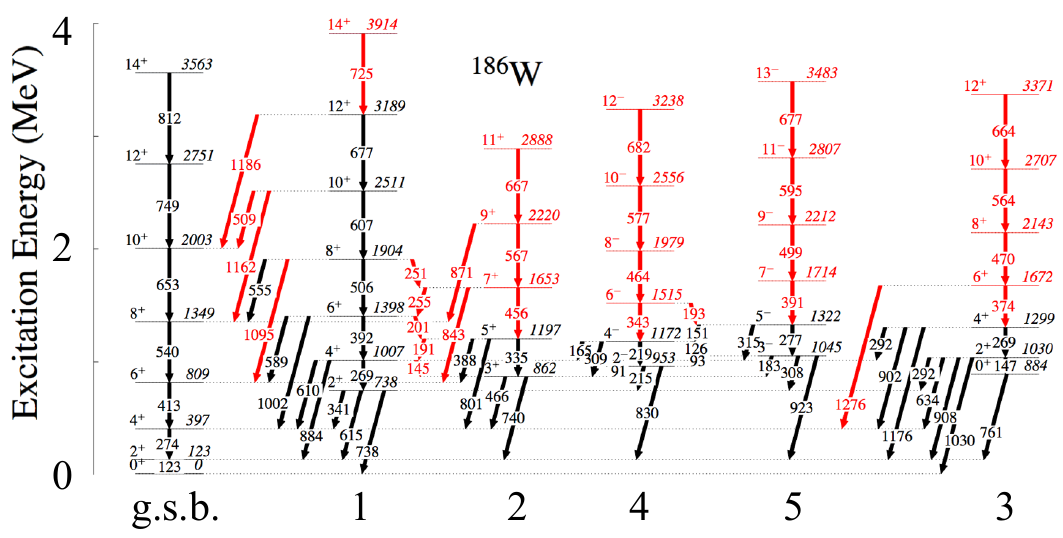} 
\caption{Partial level scheme of $^{186}$W deduced from the present work with the ground-state band (g.s.b.); $K=2^+$, $\gamma$ band (1 and 2); $K=0^+$ band (3); and $K=2^-$, octupole band (4 and 5). The new levels and transitions found in this work are indicated in red (light gray).}
 \label{fig4}
\end{center}
\end{figure*}

The proposed expansion of the level scheme is presented in Fig.~\ref{fig4}. Several rotational bands established in prior studies of $^{186}$W (e.g. Ref.~\cite{kulessa1989}) provide the foundation for the present work. Assignments of new transitions to these bands were based on observed double- and triple-coincidence relationships with transitions between established levels. A summary of the observed excited states in $^{186}$W, and of the $\gamma$-ray transitions connecting them, is provided in Table~\ref{table1}. The ground-state band (g.s.b.) was confirmed up to its $J^{\pi}=14^{+}$ member. Several side bands were observed: the $K^{\pi}=2^+$, $\gamma$ band (labeled 1 and 2 in Fig.~\ref{fig4}); the $K^{\pi}=0^+$ band (3); and the $K^{\pi}=2^-$, octupole band (4 and 5). These were extended up to $J^{\pi} = 14^{+}$, $J^{\pi} = 12^{+}$, and $J^{\pi} = 13^{-}$, respectively. 

Relative $\gamma$-ray branching ratios have been deduced by gating on a single transition feeding into the state under consideration. Yields of the observed depopulating $\gamma$~rays were corrected for their relative detection efficiencies and normalized to the strongest transition from each level. There was good agreement with values known from the literature \cite{monnand1969}.

Directional correlations of $\gamma$ rays emitted from oriented states - the `DCO ratio method' \cite{kramerflecken1989} - were used to determine multipolarities of isolated transitions, whenever the measured statistics allowed. Two angle-dependent, asymmetric, $\gamma$-$\gamma$ coincidence matrices were constructed from the sorted data. One axis on each matrix had no angle restrictions on the detected $\gamma$ ray, while the other axis was restricted to only include gamma rays detected within a limited angular range with respect to the direction of motion of the recoiling nucleus. 

While conventional DCO ratios utilize angles measured with respect to the beam direction, anisotropies with respect to the recoil direction of the nucleus are expected to be comparatively enhanced. The experimental DCO ratios ($R_{\rm{DCO}}$) were calculated using $ \gamma $-ray intensities extracted from angle-constrained spectra such that: \\[-0.4cm]

\begin{equation}
{R}_{\rm{DCO}(\gamma)} = \frac{{I}_{\gamma}(0^{\circ} - 20^{\circ})}{{I}_{\gamma}(80^{\circ} - 100^{\circ})}, \\[0.2cm]
\end{equation}

\noindent 
where ${I}_{\gamma}(0^{\circ} - 20^{\circ})$ and ${I}_{\gamma}(80^{\circ} - 100^{\circ})$ correspond to the measured intensities of $ \gamma $ rays emitted within the angle ranges of $ 0^{\circ} - 20^{\circ} $ and $ 80^{\circ} - 100^{\circ} $, respectively. The ratio values were normalized to the measured ratio for the 6$^{+}_{1} \to 4^{+}_{1}$, stretched quadrupole transition, which was found by gating on the 4$^{+}_{1} \to 2^{+}_{1}$, $\gamma$ ray. With this prescription, stretched quadrupole transitions were found to cluster around $R_{\rm{DCO}} \approx 1$, and stretched dipole ones were around $R_{\rm{DCO}} \approx 0.5$. The measured values, provided in


\clearpage
\setlength{\LTcapwidth}{17.5cm}
\begin{longtable*}{@{\extracolsep{0.6cm}}lccccccccc}
\caption{Summary of the $\gamma$-ray transitions and excited states in $^{186}$W observed in this work. Initial-level (${E_i}$), final-level (${E_f}$), and $\gamma$-ray (${E}_{\gamma}$) energies are given in keV; uncertainties are discussed in the text. Spins and parities (${J}^{\pi}_{i,f}$) and band placements ($\rm{Band}_{i,f}$) are from Ref.~\cite{A=186} or proposed from the current work. Transition-intensity branching ratios (${I}_{\gamma}$) are normalized to the strongest transition depopulating each level [100 units]. Measured DCO ratios (${R}_{\rm{DCO}}$) and transition multipolarities ($\sigma L$) are provided and discussed in the text; assignments marked with * are from Ref.~\cite{A=186}, while dipole ($D$) and quadrupole ($Q$) assignments marked $\dagger$ are proposed from the ${R}_{\rm{DCO}}$ values of this work.} \\

\hline\hline \\[-0.2cm]
\multicolumn{1}{l}{${E_i}$} & 
\multicolumn{1}{c}{${J}^{\pi}_i$} & 
\multicolumn{1}{c}{$\rm{Band}_i$} & 
\multicolumn{1}{c}{${E}_{\gamma}$} & 
\multicolumn{1}{c}{${I}_{\gamma}$} & 
\multicolumn{1}{c}{${R}_{\rm{DCO}}$}   &
\multicolumn{1}{c}{$\sigma L$} & 
\multicolumn{1}{c}{${E_f}$} & 
\multicolumn{1}{c}{${J}^{\pi}_f$} & 
\multicolumn{1}{c}{$\rm{Band}_f$} \\[0.1cm] 
\multicolumn{1}{l}{(keV)} & 
\multicolumn{1}{c}{} & 
\multicolumn{1}{c}{} & 
\multicolumn{1}{c}{(keV)} & 
\multicolumn{1}{c}{($\%$)} & 
\multicolumn{1}{c}{} & 
\multicolumn{1}{c}{} & 
\multicolumn{1}{c}{(keV)} & 
\multicolumn{1}{c}{} & 
\multicolumn{1}{c}{}   \\[0.1cm]
\hline\hline  \\[-0.2cm]
\endfirsthead

\multicolumn{10}{c}%
{{\tablename\ \thetable{} -- continued}} \\
\hline\hline \\[-0.2cm]
\multicolumn{1}{l}{${E_i}$} & 
\multicolumn{1}{c}{${J}^{\pi}_i$} & 
\multicolumn{1}{c}{$\rm{Band}_i$} & 
\multicolumn{1}{c}{${E}_{\gamma}$} & 
\multicolumn{1}{c}{${I}_{\gamma}$} & 
\multicolumn{1}{c}{${R}_{\rm{DCO}}$}   &
\multicolumn{1}{c}{$\sigma L$} & 
\multicolumn{1}{c}{${E_f}$} & 
\multicolumn{1}{c}{${J}^{\pi}_f$} & 
\multicolumn{1}{c}{$\rm{Band}_f$} \\[0.1cm] 
\multicolumn{1}{l}{(keV)} & 
\multicolumn{1}{c}{} & 
\multicolumn{1}{c}{} & 
\multicolumn{1}{c}{(keV)} & 
\multicolumn{1}{c}{($\%$)} & 
\multicolumn{1}{c}{} & 
\multicolumn{1}{c}{} & 
\multicolumn{1}{c}{(keV)} & 
\multicolumn{1}{c}{} & 
\multicolumn{1}{c}{}   \\[0.1cm]
\hline\hline  \\[-0.2cm]
\endhead

 \\[-0.55cm]
\hline\hline  
\endfoot

0.0	        &	0$^{+}_{\rm{1}}$	&	g.s.b.	&	--	   	&	--		&	--		&	--		&	--		&	--				&	--	\\ [0.05cm] \\[-0.1cm]
122.64(2)	&	2$^{+}_{\rm{1}}$	&	g.s.b.	&	122.6	&	100		&	0.925(5)	&	$E2$*	&	0		&	0$^{+}_{\rm{1}}$	&	g.s.b.	\\ [0.05cm] \\[-0.1cm]
396.6(1)	&	4$^{+}_{\rm{1}}$	&	g.s.b.	&	273.9	&	100		&	1.006(6)	&	$E2$*	&	122.64(2)	&	2$^{+}_{\rm{1}}$	&	g.s.b.	\\ [0.05cm] \\[-0.1cm]
737.8(3)	&	2$^{+}_{\rm{2}}$	&	1	&	341.0	&	0.9(1)	&			&	$[E2]$*	&	396.6(1)	&	4$^{+}_{\rm{1}}$	&	g.s.b.	\\ [0.1cm]
        		&	               		 	&		&	615.3	&	95.6(29)	&			& $M1+E2$*	&	122.64(2)	&	2$^{+}_{\rm{1}}$	&	g.s.b.	\\ [0.1cm]
      	  	&	   		             	&		&	738.0	&	100		&			&	$E2$*	&	0		&	0$^{+}_{\rm{1}}$	&	g.s.b.	\\ [0.05cm] \\[-0.1cm]
809.3(1)	&	6$^{+}_{\rm{1}}$	&	g.s.b.	&	412.7	&	100		&	1.000(4)	&	$E2$*	&	396.6(1)	&	4$^{+}_{\rm{1}}$	&	g.s.b.	\\ [0.05cm] \\[-0.1cm]
862.3(1)	&	3$^{+}_{\rm{1}}$	&	2	&	465.7	&	32.7(11)	&			& $(M1+E2$)*	&	396.6(1)	&	4$^{+}_{\rm{1}}$	&	g.s.b.	\\ [0.1cm]
     	   	&	  		              	&		&	739.7	&	100		&	0.797(24)	& $(M1+E2$)*	&	122.64(2)	&	2$^{+}_{\rm{1}}$	&	g.s.b.	\\ [0.05cm] \\[-0.1cm]
883.60(3)	&	0$^{+}_{\rm{2}}$	&	3	&	761.0	&	100		&			&			&	122.64(2)	&	2$^{+}_{\rm{1}}$	&	g.s.b.	\\ [0.05cm] \\[-0.1cm]
952.7(1)	&	2$^{-}_{\rm{1}}$	&	4	&	90.6(1)	&	20.3(11)	&			&	($E1$)*	&	862.3(1)	&	3$^{+}_{\rm{1}}$	&	2	\\ [0.1cm]
    	    	&	  		              	&		&	214.8	&	100		&			&	$E1$*	&	737.8(3)	&	2$^{+}_{\rm{2}}$	&	1	\\ [0.1cm]
    	    	&	   		             	&		&	830.1	&	3.3(5)	&			& ($E1+M2$)*	&	122.64(2)	&	2$^{+}_{\rm{1}}$	&	g.s.b.	\\ [0.05cm] \\[-0.1cm]
1006.7(1)	&	4$^{+}_{\rm{2}}$	&	1	&	144.5(1)	&	0.7(1)	&			&			&	862.3(1)	&	3$^{+}_{\rm{1}}$	&	2	\\ [0.1cm]
		&					&		&	268.9	&	6.3(2)	&			&	$[E2]$*	&	737.8(3)	&	2$^{+}_{\rm{2}}$	&	1	\\ [0.1cm]
    	    	&	   		             	&		&	610.2	&	100		&			& ($M1+E2$)*   &	396.6(1)	&	4$^{+}_{\rm{1}}$	&	g.s.b.	\\ [0.1cm]
    	    	&	   		             	&		&	884.1	&	58.7(17)	&	1.204(42)	&	$E2$*	&	122.64(2)	&	2$^{+}_{\rm{1}}$	&	g.s.b.	\\ [0.05cm] \\[-0.1cm]
1030.2(6)	&	2$^{+}_{\rm{3}}$	&	3	&	146.6(1)	&	$<$3		&			&			&	883.60(3)	&	0$^{+}_{\rm{2}}$	&	3	\\ [0.1cm]
    	    	&	    		            	&		&	292.4(5)	&	14.4(9)	&			&			&	737.8(3)	&	2$^{+}_{\rm{2}}$	&	1	\\ [0.1cm]
    	    	&	    		            	&		&	633.7	&	58.8(29)	&			&	$Q$*		&	396.6(1)	&	4$^{+}_{\rm{1}}$	&	g.s.b.	\\ [0.1cm]
     	   	&	    		            	&		&	907.6	&	100		&			& ($M1+E2$)*	&	122.64(2)	&	2$^{+}_{\rm{1}}$	&	g.s.b.	\\ [0.1cm]
     	   	&	    		            	&		&	1030.2	&	67.7(24)	&			&	$E2$*	&	0		&	0$^{+}_{\rm{1}}$	&	g.s.b.	\\ [0.05cm] \\[-0.1cm]
1045.4(5)	&	3$^{-}_{\rm{1}}$	&	5	&	92.7		&	$<$3 	&			&  $M1+E2$*	&	952.7(1)	&	2$^{-}_{\rm{1}}$	&	4	\\ [0.1cm]
     	   	&	  		              	&		&	183.1	&	32.1(10)	&			&	$E1$*	&	862.3(1)	&	3$^{+}_{\rm{1}}$	&	2	\\ [0.1cm]
     	   	&	  		              	&		&	307.5	&	100		&			&	$E1$*	&	737.8(3)	&	2$^{+}_{\rm{2}}$	&	1	\\ [0.1cm]
     	   	&	 		               	&		&	922.8	&	9.5(5)	&			&			&	122.64(2)	&	2$^{+}_{\rm{1}}$	&	g.s.b.	\\ [0.1cm]
    	    	&	 		               	&		&	1045		&	$<$3		&			&	$[E3]$*	&	0		&	0$^{+}_{\rm{1}}$	&	g.s.b.	\\ [0.05cm] \\[-0.1cm]
1171.6(1)	&	4$^{-}_{\rm{1}}$	&	4	&	126.3	&	$<$4		&			&			&	1045.4(5)	&	3$^{-}_{\rm{1}}$	&	5	\\ [0.1cm]
   	     	&	 		               	&		&	164.8	&	10.1(5)	&			&			&	1006.7(1)	&	4$^{+}_{\rm{2}}$	&	1	\\ [0.1cm]
   	     	&	 		               	&		&	218.9	&	35.5(14)	&			&			&	952.7(1)	&	2$^{-}_{\rm{1}}$	&	4	\\ [0.1cm]
    	    	&	 		               	&		&	309.4	&	100		&			&	$D(+Q)$*	&	862.3(1)	&	3$^{+}_{\rm{1}}$	&	2	\\ [0.05cm] \\[-0.1cm]
1197.4(1)	&	5$^{+}_{\rm{1}}$	&	2	&	190.6(1)	&	$<$1		&			&			&	1006.8(1)	&	4$^{+}_{\rm{2}}$	&	1	\\ [0.1cm]
		&					&		&	335.0	&	30.3(10)	&			&	$Q$*		&	862.3(1)	&	3$^{+}_{\rm{1}}$	&	2	\\ [0.1cm]
     	   	&	 		               	&		&	388.2	&	3.5(2)	&			&			&	809.3(1)	&	6$^{+}_{\rm{1}}$	&	g.s.b.	\\ [0.1cm]
    	    	&				         &		&	800.7	&	100		&	0.634(13)	&	$D+Q$*    &	396.6(1)	&	4$^{+}_{\rm{1}}$	&	g.s.b.	\\ [0.35cm]
1298.9(2)	&	4$^{+}_{\rm{3}}$	&	3	&	268.5(2)	&	72.4(29)	&			&	$[E2]$*	&	1030.2(6)	&	2$^{+}_{\rm{3}}$	&	3	\\ [0.1cm]
    	    	&	 		               	&		&	292.2(5)	&	7.1(6)	&			&			&	1006.7(1)	&	4$^{+}_{\rm{2}}$	&	1	\\ [0.1cm]
    	    	&	 		               	&		&	902.4	&	73.2(33)	&	0.900(78)	&  	$D+Q$*	&	396.6(1)	&	4$^{+}_{\rm{1}}$	&	g.s.b.	\\ [0.1cm]
     	   	&	   		             	&		&	1176.3	&	100		&			&	$E2$*	&	122.64(2)	&	2$^{+}_{\rm{1}}$	&	g.s.b.	\\ [0.05cm] \\[-0.1cm]
1322.1(2)	&	5$^{-}_{\rm{1}}$	&	5  	&	150.5(1)	&	9.9(4)	&			&			&	1171.6(1)	&	4$^{-}_{\rm{1}}$	&	4	\\ [0.1cm]
      	  	&	   		             	&		&	276.7  	&	100		&			&			&	1045.4(5)	&	3$^{-}_{\rm{1}}$	&	5	\\ [0.1cm]
     	   	&	   		             	&		&	315.4  	&	76.6(36)	&	0.638(61)	&	$D(+Q)$*	&	1006.7(1)	&	4$^{+}_{\rm{2}}$	&	1	\\ [0.05cm] \\[-0.1cm]
1349.3(1)	&	8$^{+}_{\rm{1}}$	&	g.s.b.	&	540.0  	&	100		&	1.077(6)	&	$E2$*	&	809.3(1)	&	6$^{+}_{\rm{1}}$	&	g.s.b.	\\ [0.05cm] \\[-0.1cm]
1398.1(1)	&	6$^{+}_{\rm{2}}$	&	1	&	200.7(1)  	&	5.2(2)	&			&			&	1197.4(1)	&	5$^{+}_{\rm{1}}$	&	2	\\ [0.1cm]
		&					&		&	391.5  	&	100		&			&	$Q$*		&	1006.7(1)	&	4$^{+}_{\rm{2}}$	&	1	\\ [0.1cm]
     	   	&	  		              	&		&	588.7  	&	69.3(21)	&	0.493(12)	&  	$D^{\dagger}$ &	809.3(1)	&	6$^{+}_{\rm{1}}$	&	g.s.b.	\\ [0.1cm]
     	   	&	  		              	&		&	1001.6 	&	54.3(17)	&	0.995(27)	&	$Q$*		&	396.6(1)	&	4$^{+}_{\rm{1}}$	&	g.s.b.	\\ [0.05cm] \\[-0.1cm]
1514.7(2)	&	6$^{-}_{\rm{1}}$	&	4	&	192.5(1)	&	$<$5		&			&			&	1322.2(1)&	5$^{-}_{\rm{1}}$	&	5	 \\[0.1cm]
		&					&		&	343.0(2)	&	100		&			&			&	1171.6(1)	&	4$^{-}_{\rm{1}}$	&	4	\\ [0.05cm] \\[-0.1cm]
1652.8(2)	&	7$^{+}_{\rm{1}}$	&	2	&	254.6(1)	&	$<$1		&			&			&	1398.1(1)	&	6$^{+}_{\rm{2}}$	&	1	\\ [0.1cm]
		&					&		&	455.6(2)	&	100		&	0.925(37)	&	$Q$*		&	1197.4(1)	&	5$^{+}_{\rm{1}}$	&	2	\\ [0.1cm]
          	&	                 	 	&		&	843.4(3)	&	49.4(23)	&	0.479(29)	&  	$D$*		&	809.3(1)	&	6$^{+}_{\rm{1}}$	&	g.s.b.	\\ [0.05cm] \\[-0.1cm]
1672.4(2)	&	6$^{+}_{\rm{3}}$	&	3	&	373.6(2)	&	100		&			&			&	1298.9(2)	&	4$^{+}_{\rm{3}}$	&	3	\\ [0.1cm]
     	   	&	    		            	&		&	1275.7(3)	&	66.5(34)	&			&			&	396.6(1)	&	4$^{+}_{\rm{1}}$	&	g.s.b.	\\ [0.05cm] \\[-0.1cm]
1713.6(2)	&	7$^{-}_{\rm{1}}$	&	5	&	391.4(2)	&	100		&			&			&	1322.1(2)	&	5$^{-}_{\rm{1}}$	&	5	\\ [0.05cm] \\[-0.1cm]
1904.0(1)	&	8$^{+}_{\rm{2}}$	&	1	&	251.2(1)	&	$<$1		&			&			&	1652.8(2)	&	7$^{+}_{\rm{1}}$	&	2	\\ [0.1cm]
		&					&		&	506.1(2)	&	100		&	0.946(18)	&	$Q^{\dagger}$	 	&	1398.1(1)	&	6$^{+}_{\rm{2}}$	&	1	\\ [0.1cm]
    	    	&	    		            	&		&	554.9(2)	&	6.6(2)	&	0.442(24)	&	$D^{\dagger}$		&	1349.3(1)	&	8$^{+}_{\rm{1}}$	&	g.s.b.	\\ [0.1cm]
    	    	&	    		            	&		&	1094.5(3)	&	5.0(2)	&	1.005(67)	&	$Q^{\dagger}$		&	809.3(1)	&	6$^{+}_{\rm{1}}$	&	g.s.b.	\\ [0.05cm] \\[-0.1cm]
1979.0(3)	&	8$^{-}_{\rm{1}}$	&	4	&	464.4(2)	&	100		&			&			&	1514.7(2)	&	6$^{-}_{\rm{1}}$	&	4	\\ [0.05cm] \\[-0.1cm]
2002.5(2)	&	10$^{+}_{\rm{1}}$	&	g.s.b.	&	653.2   	&	100		&	1.003(10)	&	$E2$*	&	1349.3(1)	&	8$^{+}_{\rm{1}}$	&	g.s.b.	\\ [0.05cm] \\[-0.1cm]
2142.7(3)	&	8$^{+}_{\rm{3}}$	&	3	&	470.3(2)	&	100		&			&			&	1672.4(2)	&	6$^{+}_{\rm{3}}$	&	3	\\ [0.05cm] \\[-0.1cm]
2212.0(3)	&	9$^{-}_{\rm{1}}$	&	5	&	498.5(2)	&	100		&			&			&	1713.6(2)	&	7$^{-}_{\rm{1}}$	&	5	\\ [0.05cm] \\[-0.1cm]
2220.3(2)	&	9$^{+}_{\rm{1}}$	&	2	&	567.3(2)	&	100		&	1.129(87)	&	$Q^{\dagger}$		&	1652.8(2)	&	7$^{+}_{\rm{1}}$	&	2	\\ [0.1cm]
      	  	&	     		           	&		&	871.2(3)	&	14.9(37)	&			&			&	1349.3(1)	&	8$^{+}_{\rm{1}}$	&	g.s.b.	\\ [0.15cm] \\[0.1cm]
2511.3(2)	&	10$^{+}_{\rm{2}}$	&	1	&	509.1(2)	&	14.1(18)	&	0.552(22)	&	$D^{\dagger}$		 &	2002.5(2)	&	10$^{+}_{\rm{1}}$	&	g.s.b.	\\ [0.1cm]
    	    	&	      		          	&		&	607.1(2)	&	100		&	1.163(55)	&	$Q^{\dagger}$		&	1904.0(1)	&	8$^{+}_{\rm{2}}$	&	1	\\ [0.1cm]
    	    	&				         &		&	1161.9(3)	&	$<$4		&			&			&	1349.3(1)	&	8$^{+}_{\rm{1}}$	&	g.s.b.	\\ [0.05cm] \\[-0.1cm]
2555.8(4)	&	10$^{-}_{\rm{1}}$	&	4	&	576.8(2)	&	100		&			&			&	1979.0(3)	&	8$^{-}_{\rm{1}}$	&	4	\\ [0.05cm] \\[-0.1cm]
2707.0(4)	&	10$^{+}_{\rm{3}}$	&	3	&	564.4(2)	&	100		&			&			&	2142.7(3)	&	8$^{+}$			&	3	\\ [0.05cm] \\[-0.1cm]
2751.0(3)	&	12$^{+}_{\rm{1}}$	&	g.s.b.	&	748.5(2)   	&	100		&	1.002(18)	&	$Q^{\dagger}$		&	2002.5(2)	&	10$^{+}_{\rm{1}}$	&	g.s.b.	\\ [0.05cm] \\[-0.1cm]
2806.5(4)	&	11$^{-}_{\rm{1}}$	&	5	&	594.5(2)	&	100		&			&			&	2212.0(3)	&	9$^{-}_{\rm{1}}$	&	5	\\ [0.40cm]
2887.5(3)	&	11$^{+}_{\rm{1}}$	&	2	&	667.2(2)	&	100		&			&			&	2220.3(2)	&	9$^{+}$			&	2	\\ [0.05cm] \\[-0.1cm]
3188.6(2)	&	12$^{+}_{\rm{2}}$	&	1	&	677.1(2)	&	100		&			&			&	2511.3(2)	&	10$^{+}$			&	1	\\ [0.1cm]
     	   	&				        &		&	1186.3(3)	&	$<$20	&			&			&	2002.5(2)	&	10$^{+}_{\rm{1}}$	&	g.s.b.	\\ [0.05cm] \\[-0.1cm]
3237.8(4)	&	12$^{-}_{\rm{1}}$	&	4	&	682.0(2)	&	100		&			&			&	2555.8(4)	&	10$^{-}_{\rm{1}}$	&	4	\\ [0.05cm] \\[-0.1cm]
3371.1(4)	&	12$^{+}_{\rm{3}}$	&	3	&	664.1(2)	&	100		&			&			&	2707.0(4)	&	10$^{+}$			&	3	\\ [0.05cm] \\[-0.1cm]
3483.3(4)	&	13$^{-}_{\rm{1}}$	&	5	&	676.8(2)	&	100		&			&			&	2806.5(4)	&	11$^{-}_{\rm{1}}$	&	5	\\ [0.05cm] \\[-0.1cm]
3562.5(4)	&	14$^{+}_{\rm{1}}$	&	g.s.b.	&	811.5(3)   	&	100		&	1.073(45)	&	$Q^{\dagger}$		&	2751.0(2)	&	12$^{+}_{\rm{1}}$	&	g.s.b.	\\ [0.05cm] \\[-0.1cm]
3913.6(3)	&	14$^{+}_{\rm{2}}$	&	1	&	725.1(2)	&	100		&			&			&	3188.6(2)	&	12$^{+}$			&	1	\\[-0.1cm]

\label{table1}
\end{longtable*}

\noindent
Table~\ref{table1}, were consistent with adopted multipolarity assignments for known transitions in $^{186}$W  \cite{A=186}. Tentative assignments, those suggested from this work, are also included.  

In the table, energies of the well-known transitions are taken from Ref.~\cite{A=186}; these are used as a primary calibration source and are given without uncertainties, as our current measurements do not improve them. All of the new transitions, and their inferred level energies, are presented with their associated uncertainties. The dynamically reconstructed spectra strongly depend on precise measurement of the emission angles and velocities of the decaying nuclei, so the energy resolution is approximately five times poorer than for static source measurements. 

All of the new transitions are associated with known bands, most of which already have firmly assigned spins and parities. As the new states are extensions of these bands, and those show almost constant moments of inertia, there is high confidence in the level assignments presented in Table~\ref{table1}, which correspond to simple rotation of the mean field with its associated cascades of quadrupole decays. The measured DCO ratios support the proposed assignments. Consequently, we consider all new spin and parity assignments to be `firm' in the subsequent analysis.

\begin{figure}[hb!]
   \centering
   \includegraphics*[trim=0cm 5.5cm 0cm 0cm, clip=true, width=8.5cm]{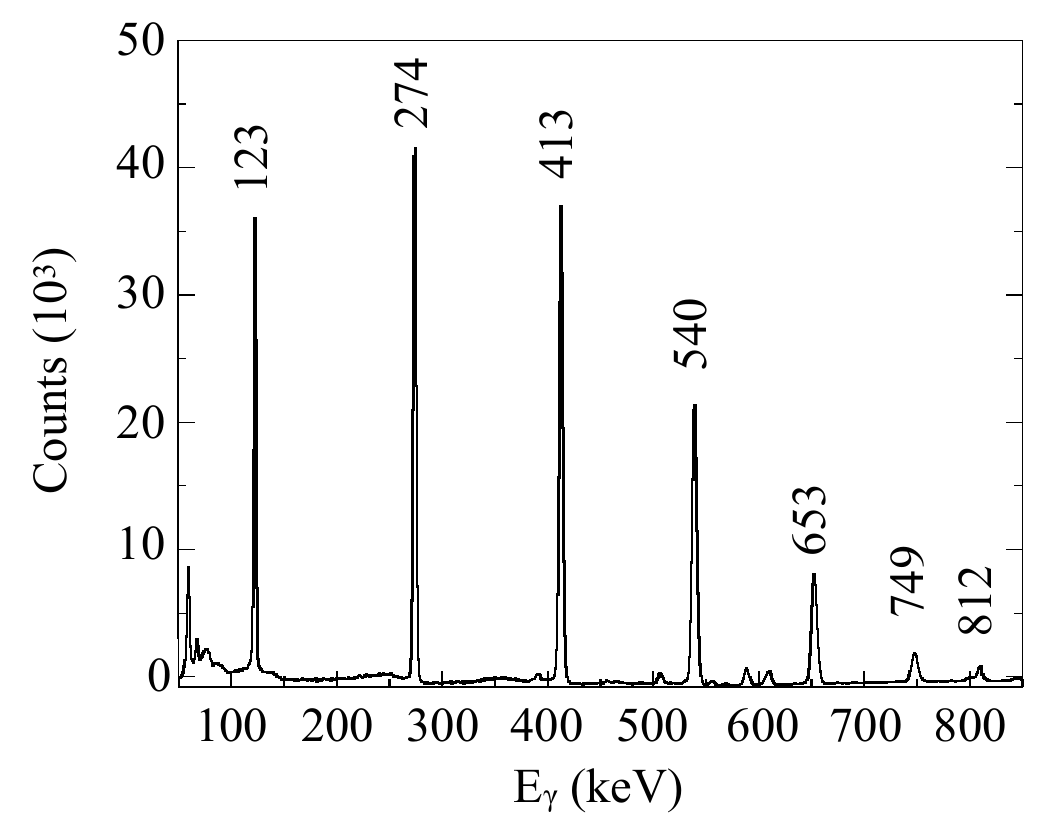} 
   \caption{Triples coincidence spectrum with a sum of double-gate combinations of all pairs of known transitions in the $^{186}$W ground-state band; intraband transition energies are labeled.}
   \label{fig5}
\end{figure}

\subsection{THE GROUND-STATE BAND}
A triples coincidence spectrum with a sum of double-gate combinations of all pairs of known transitions in the $^{186}$W ground-state band is presented in Fig.~\ref{fig5}. No additional transitions were observed that would extend this band beyond the known $J^{\pi} = 14^{+}_{\rm{1}}$ state. This was surprising, as a similar, previous study of $^{180}$Hf \cite{ngijoi-yogo2007,tandel2008-1} populated states beyond $J^{\pi} = 20^{+}$. Two factors are thought to have contributed to this outcome. 

\begin{figure}[b!]
   \centering
   \includegraphics*[trim=0cm 5.5cm 0cm 0cm, clip=true, width=8.5cm]{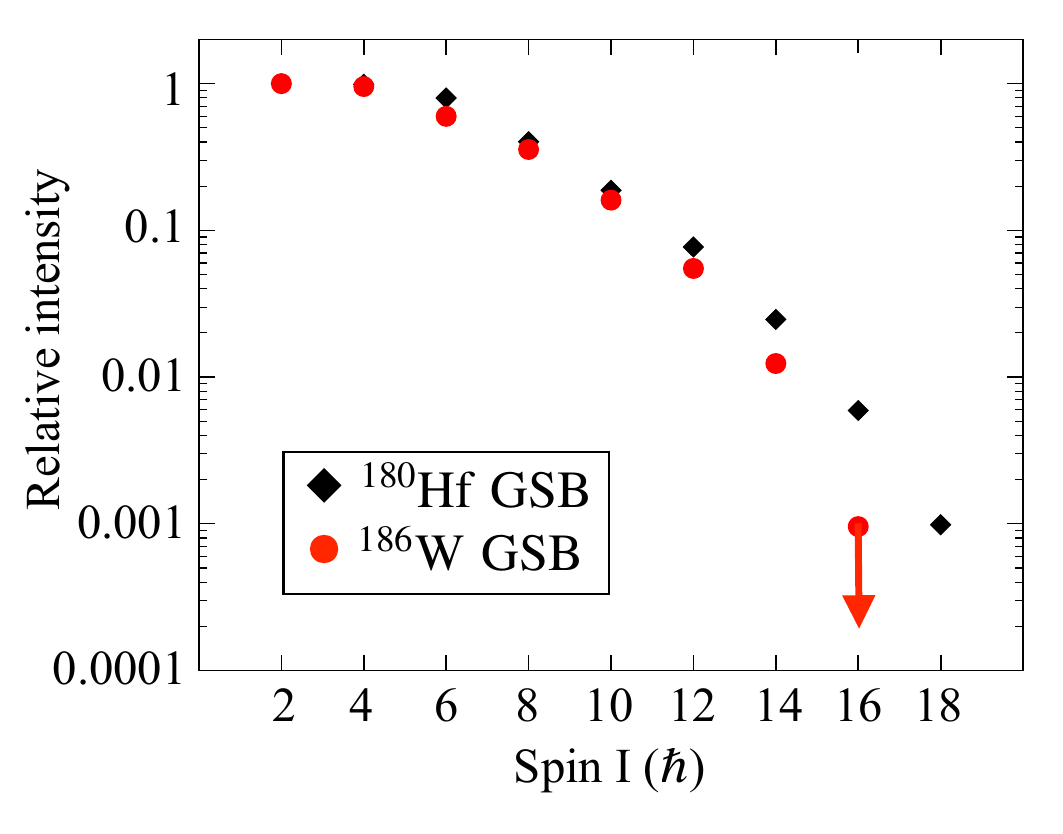} 
   \caption{Relative yields of ground-state band transitions for $^{186}$W from this work (red circles) and $^{180}$Hf from Ref.~\cite{ngijoi-yogo2007} (black diamonds) using $^{136}$Xe beams at similar energies. The yields are normalized to the 2$^{+} \to 0^{+}$ transition in each nucleus; experimental uncertainties of $\approx 1\% $ are smaller than the drawn data points.}
   \label{fig6}
\end{figure}

Most significantly, the current data set was more than an order of magnitude smaller than in the hafnium work, due to technical difficulties experienced during the experiment. Beyond that, despite near-identical experimental conditions, relative intensities of $\gamma$ rays in the $^{186}$W ground-state band decrease faster with spin than in $^{180}$Hf; this is illustrated in Fig.~\ref{fig6} and the data are available in Table~\ref{table2}.

\begin{table}[h]
\begin{center}
\caption{\label{table2} Relative yields of ground-state band transitions for $^{186}$W from the present work, normalized to the $2^+ \to 0^+$ transition; statistical uncertainties are $\approx 1\%$.}
\vspace{0.2cm}
\renewcommand{\arraystretch}{1.5}
\begin{tabular*}{\columnwidth}{@{\extracolsep{\fill}} c r}
\hline \hline
Spin, $I$ ($\hbar$) & $^{186}$W	 \\[0.05cm]
\hline \hline

2 &  1.00		 \\
4 &  0.96		 \\
6 &	0.60	 	 \\
8 & 	0.36		 \\
10 &	0.16	 	 \\
12 &	 0.06		 \\
14 & 	0.01	          \\
16 & 	$<$0.001		 \\

\hline
\hline
\end{tabular*}
\end{center}
\end{table}

As the moments of inertia and low-lying collectivity of these two nuclei are similar, this difference appears to arise from smaller decay matrix elements at high spin in the tungsten case. This property might be associated with a change to an oblate shape, as predicted by the calculation of Ref.~\cite{hilton1979}, but a dedicated experiment would be required to address this issue further.

\subsection{THE \boldmath{$K^{\pi} = 2^{+}$}, \boldmath{$\gamma$} BAND}

\begin{figure}[t!]
\begin{center}
\includegraphics*[trim=0cm 3.5cm 0cm 0cm, clip=true, width=8.5cm]{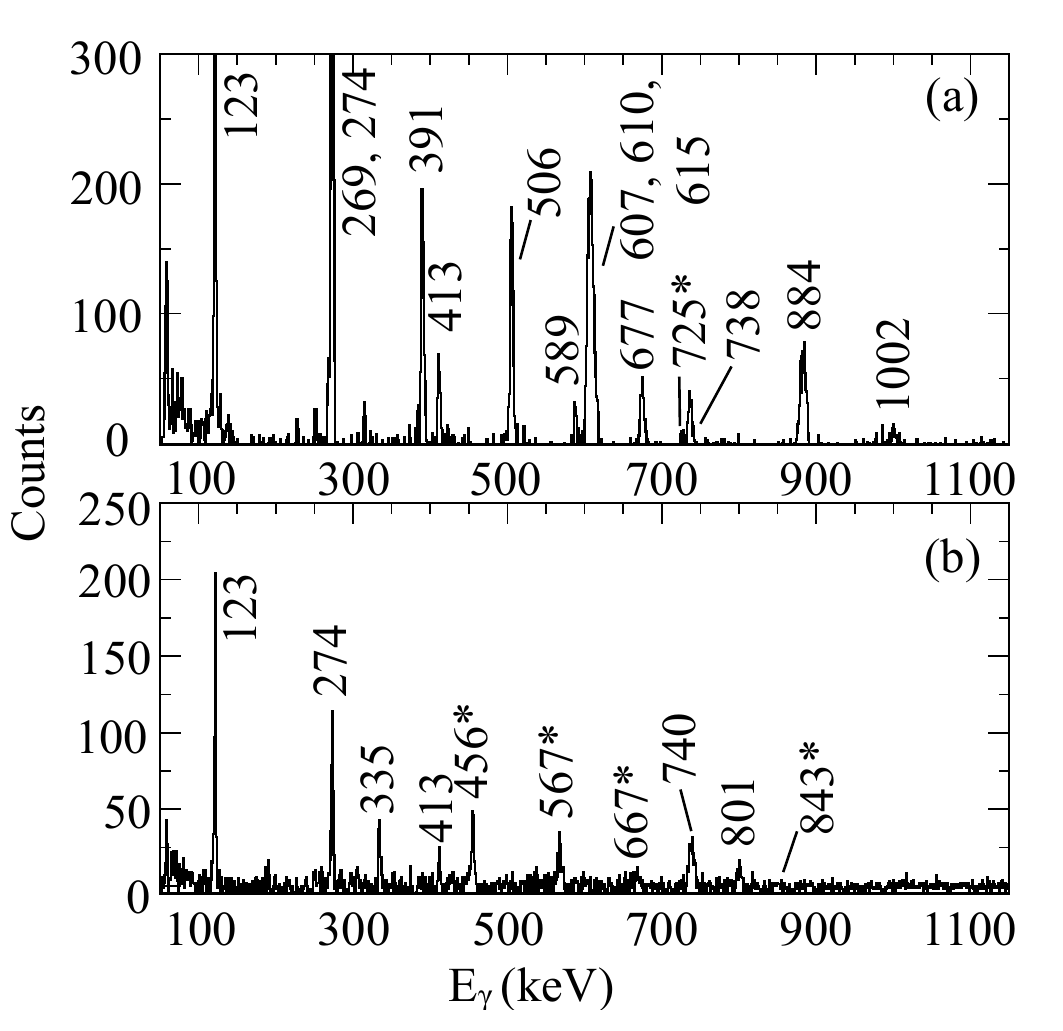} 
\caption{Summed triples coincidence spectra, double-gated on transitions in (a) band 1 and (b) band 2 of $^{186}$W. A combination of 269-, 392-, 506-, 607-, and 677-keV $\gamma$-ray energies was used for band 1. Similarly, a combination of 335-, 456-, 567-, 667-, and 740-keV $\gamma$-ray energies was used for band 2. New transitions identified in the present work are indicated with asterisks.} 
\label{fig7}
\end{center}
\end{figure}

The next-lowest collective sequence observed in $^{186}$W is the $\gamma$ vibrational band, with the bandhead located at 738~keV. Triples coincidence spectra, double-gated on transitions in bands 1 and 2, are presented in Fig.~\ref{fig7}. The ratio of $E(2^+_2)/E(2^+_1)$ excitation energies is 6.0; i.e., roughly half that seen in $^{180}$W, where the same bandhead is located at 1117~keV. From the ratio above and a rigid geometrical model \cite{davydov1958}, a triaxiality parameter of $\gamma=16.0^{\circ}$ can be deduced. However, as pointed out in Ref.~\cite{wu1996}, the deformation is likely to be dynamic and can be more reliably inferred from the distribution of electric quadrupole ($E2$) strengths between the ground-state and $\gamma$ bands, usually inferred from Coulomb-excitation measurements. For $^{186}$W, a mean value of $\gamma = 17.0(7)^{\circ}$ was inferred in Ref.~\cite{wu1996}.

\subsection{THE \boldmath{$K^{\pi} = 0^{+}$} BAND}

\begin{figure}[t!]
\begin{center}
\includegraphics*[trim=0cm 5.5cm 0cm 0cm, clip=true, width=8.5cm]{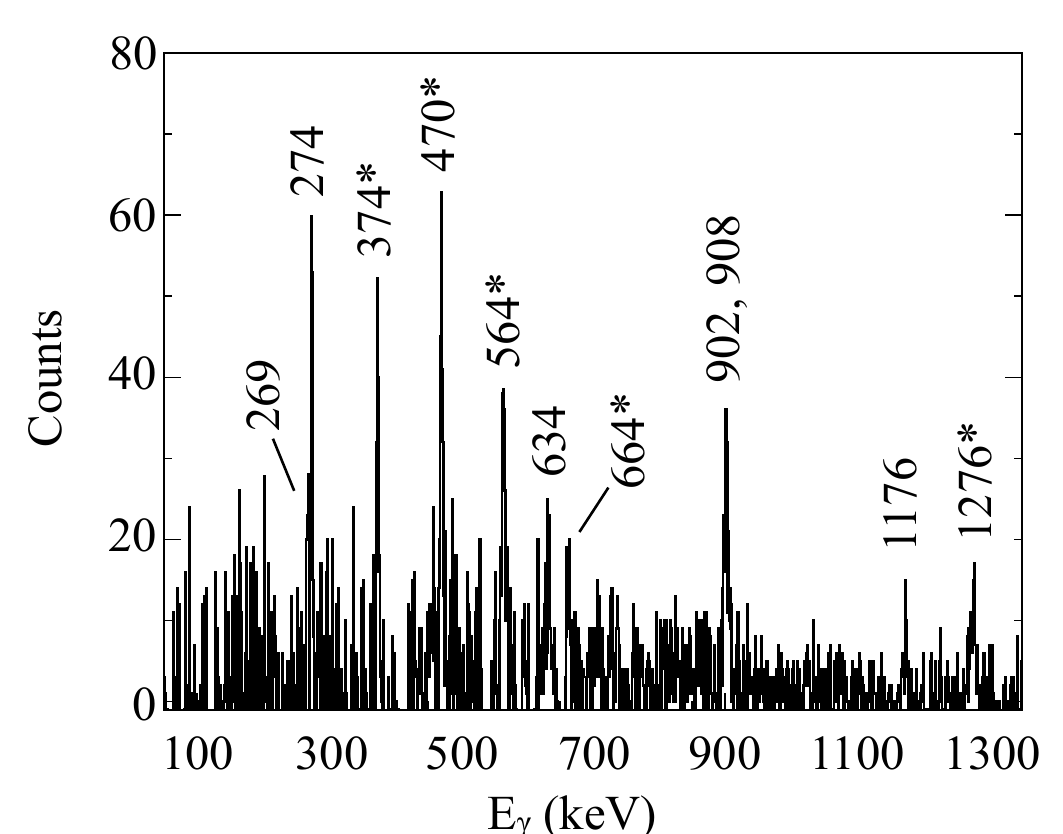} 
\caption{Summed coincidence spectrum, double-gated on transitions in band 3 of $^{186}$W. A combination of 374-, 470-, 564-, and 902-keV $\gamma$-ray energies from band 3 was used; a double gate placed on the 902-keV transition from band 3 and the 274-keV transition in the ground-state band was included to improve statistics. New transitions identified in the present work are indicated with asterisks.} 
\label{fig8}
\end{center}
\end{figure}

\begin{figure}[t!]
\begin{center}
\includegraphics*[trim=0cm 2.5cm 0cm 0cm, clip=true, width=8.5cm]{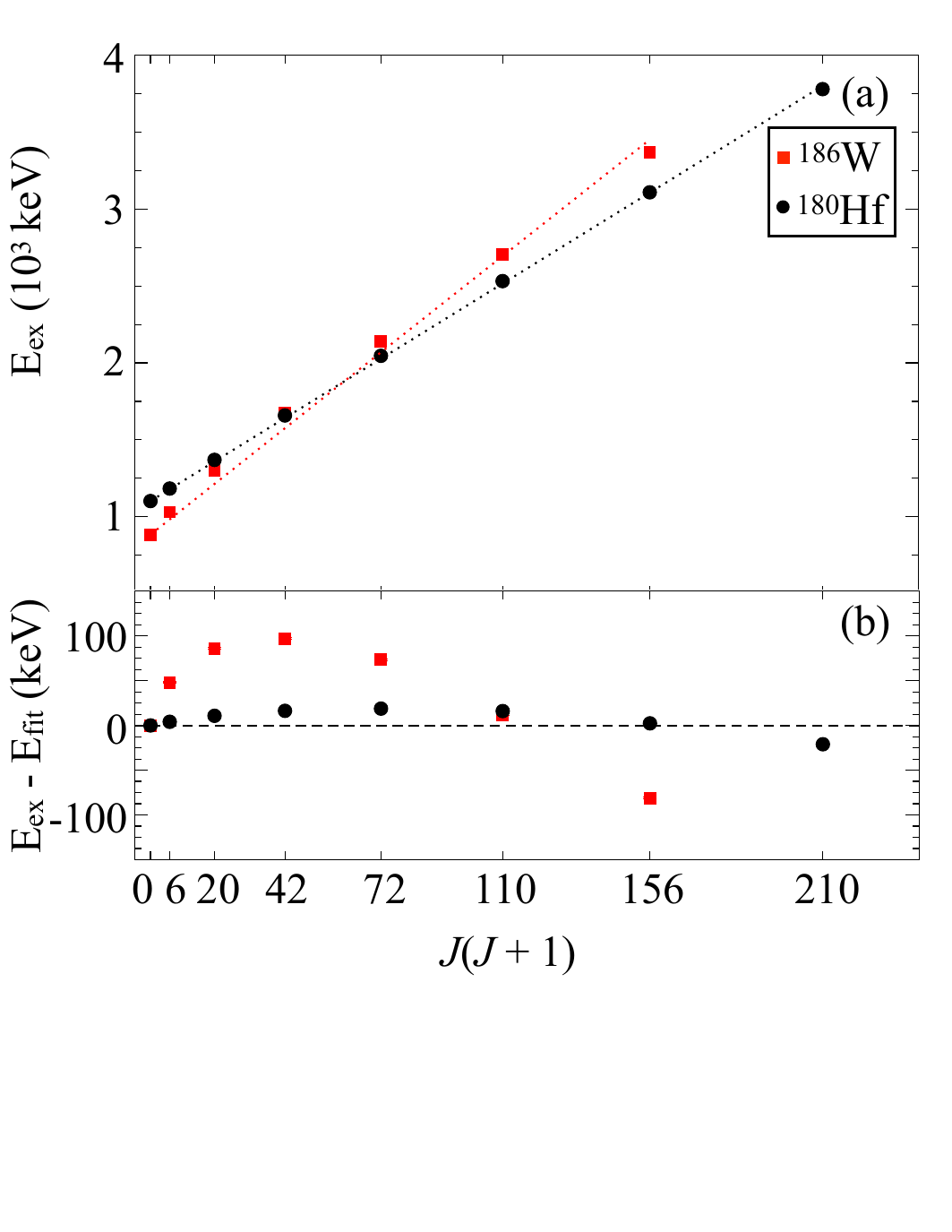} 
\caption{(a) Level energies of the $ K^{\pi} = 0^+ $ band members in $^{186}$W from the present work (red squares) and $^{180}$Hf \cite{A=180} (black circles) plotted as a function of $J(J+1)$. The dashed lines show linear lines of best fit applied to the data. The lower panel (b) shows the difference between the measured energies and the results of the linear fit for each $J(J+1)$ value. } 
\label{fig9}
\end{center}
\end{figure}

As shown in Fig.~\ref{fig4}, band 3 is built on the previously known, $J^{\pi} = 0^{+}_{\rm{2}}$ state at 884 keV, which decays to the $2^{+}_{\rm{1}}$ level by a 761-keV transition. This band was proposed to be a `quasi-$\beta$' band~\cite{milner1971} and was known up to its $4^{+}_{\rm{2}}$ member at 1299~keV. A triples coincidence spectrum, double-gated on transitions in band 3, with a double gate placed on the 902-keV transition from band 3 and the 274-keV transition in the ground-state band, is provided in Fig.~\ref{fig8}. Four new $\gamma$ rays in a cascade of 374, 470, 564, and 664~keV were added to this sequence. An additional interband transition of 1276~keV connects the $6^{+}_{\rm{3}}$ candidate to the $4^{+}_{\rm{1}}$ state. In contrast to better-known $ K^{\pi} = 0^+ $ vibrational bands in this region, such as in $^{180}$Hf \cite{A=180}, the level energies in $^{186}$W deviate from an almost perfect linear trajectory when plotted as a function of $J(J+1)$, shown in Fig.~\ref{fig9}, especially at the lowest spins. This suggests that the sequence is of more complex character.

\subsection{THE \boldmath{$K^{\pi} = 2^{-}$}, OCTUPOLE BAND}

\begin{figure}[ht!]
\begin{center}
\includegraphics*[trim=0cm 3.5cm 0cm 0cm, clip=true, width=8.5cm]{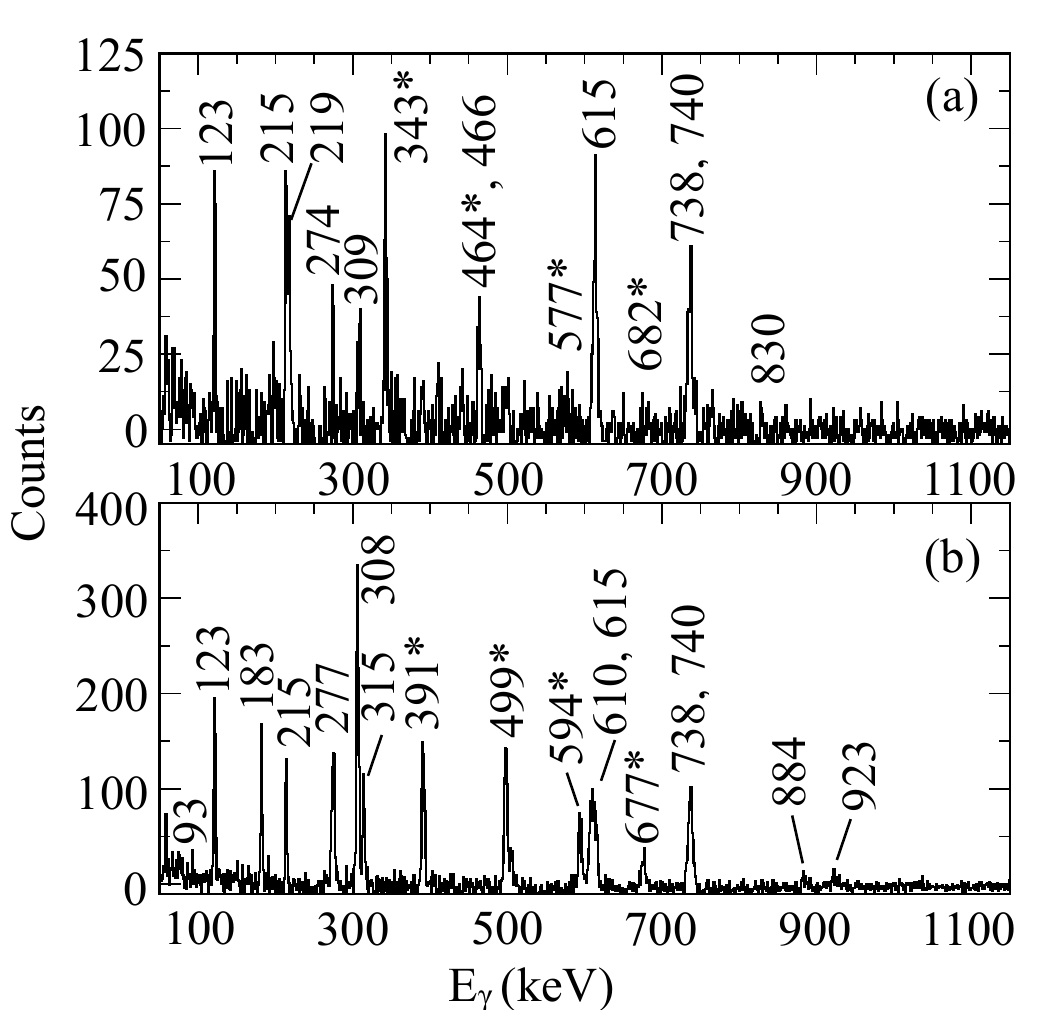} 
\caption{Summed triples coincidence spectra, double-gated on transitions in (a) band 4 and (b) band 5 of $^{186}$W. A combination of 215-, 219-, 343-, 464-, and 577-keV $\gamma$-ray energies was used for band 4. Similarly, a combination of 277-, 391-, 499-, 595-, and 677-keV $\gamma$-ray energies was used for band 5. New transitions identified in the present work are indicated with asterisks.} 
\label{fig10}
\end{center}
\end{figure}

The bandheads for bands 4 and 5 in Fig.~\ref{fig4} were previously reported with $ J^{\pi} = 2^{-} $ and $ J^{\pi} = 3^{-} $ respective assignments \cite{mcgowan1977}. Only the two lowest members of each sequence were known: the $ 2^{-}_{\rm{1}} $ (953 keV) and $ 4^{-}_{\rm{1}} $ (1172~keV) states for band 4; and the $ 3^{-}_{\rm{1}} $ (1045 keV) and $ 5^{-}_{\rm{1}} $ (1322 keV) ones for band 5. Triples coincidence spectra, double gated on transitions in bands 4 and 5, are presented in Fig.~\ref{fig10}. In this work, these bands were extended to $ J^{\pi} = 12^{-}_{\rm{1}} $ and $ J^{\pi} = 13^{-}_{\rm{1}} $, respectively. The intraband transitions are by far the strongest, although weak, interband transitions linking to the $\gamma$ band from the low-lying states were also identified. 

\section{\label{discussion}DISCUSSION}


\subsection{THE \boldmath{$K^{\pi} = 2^{+}$}, \boldmath{$\gamma$} BAND}

By finding and extending the odd-spin members of the $\gamma$ band (band 2 in Fig.~\ref{fig4}) the issue of rigid deformation versus triaxial softness can be explored through examination of the so-called even- and odd-spin staggering. A useful analysis tool for doing so was introduced in Ref.~\cite{mccutchan2007} through the expression:

\begin{equation}
S(J) = \frac{[\{E(J)-E(J-1)\} - \{E(J-1)-E(J-2)\}]}{E(2^+_{\rm{1}})}, \\[0.3cm]
\label{equ:stag}
\end{equation}

\noindent
where the staggering parameter $S(J)$ is determined from the energy differences between levels with $\Delta J = 1$ within the rotational band. This expression captures and extends the physics discussed in Ref.~\cite{zamfir1991}. The odd-spin states in a given rotational band of a rigid, deformed nucleus are expected to be located almost halfway between their even-spin neighbors. Inserting the simplest description of the level energies expected from a $K=2$, $\gamma$ band: \\[-0.6cm]

\begin{equation}
E(J) = E(0) + \frac{\hbar^2 }{ 2I } [J(J+1) - K^2], \\[0.3cm]
\label{equ:gamma}
\end{equation}

\noindent
into Eq.~\ref{equ:stag} results in a small, positive value of $S(J)=0.33$ that remains constant with increasing spin, $J$. 

In the Davydov-Filippov model of a rigid, triaxial nucleus \cite{davydov1958, moore1960}, the energies of the even-spin states are displaced upwards relative to the odd-spin band members, causing $S(J)$ values to stagger and be positive for even $J$ and small or negative for odd $J$. The amplitude of the staggering increases as the axial asymmetry parameter, $\gamma$, increases. In contrast, a $\gamma$-independent vibrational nuclear potential, such as the Wilets-Jean model \cite{wilets1956}, the even-spin states are displaced downwards relative to the odd-spin band members, small or negative $S(J)$ values occur for even-$J$ levels, and positive ones are found for even-$J$ states. If the shape and softness remain unchanged with spin, this staggering pattern should be monotonic. If its amplitude increases, changes in triaxial deformation would be increasing, either in magnitude or in softness.

\begin{figure}[ht!]
\begin{center}
\includegraphics*[trim=0cm 6cm 0cm 0cm, clip=true, width=8.5cm]{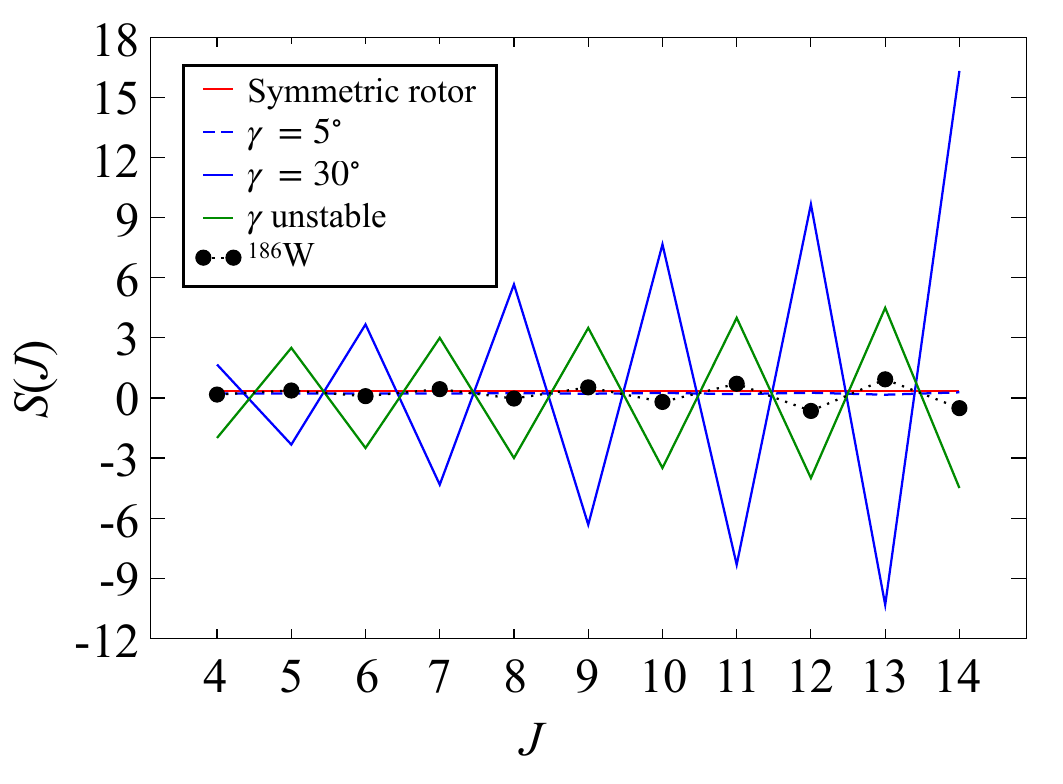} 
\caption{Predicted staggering patterns for the $K = 2^+$, $\gamma$ band in an axially symmetric rotor (red line), a rigid asymmetric rotor with $\gamma = 5^{\circ}$ (blue dashes) and $\gamma = 30^{\circ}$ (blue line), and a $\gamma$-unstable nucleus (green line). The data for $^{186}$W (black circles) are from this work.} 
\label{fig11}
\end{center}
\end{figure}

Staggering parameters predicted for each of these cases are compared to experimental values for the $K = 2^+$, $\gamma$~band in $^{186}$W, from this work, in Fig.~\ref{fig11}. As discussed above, $S(J) =  0.33$ for all $J$ in a rigid, axially symmetric rotor. A rigid, triaxial rotor with $\gamma = 5^{\circ}$, which closely resembles a symmetric nucleus, has $S(J)$ values that are small, positive and almost constant. For $\gamma = 30^{\circ}$, the large $S(J)$ values are positive for even $J$, and diverge rapidly. In contrast, the staggering pattern phase is opposite for the $\gamma$-soft model, which gives negative $S(J)$ values for even $J$ and positive ones for odd $J$. The experimental staggering pattern found in $^{186}$W has the same phase as the $\gamma$-soft model, but it is much smaller in magnitude than the Wilets-Jean limit. The degree of $\gamma$ softness is quite stable with spin, as the amplitude of the staggering increases smoothly and gradually, and it does not show evidence of any drastic change of structure, at least not below spin $J = 15$.

With this tool, the landscape of axial symmetry and softness can be explored both in ($N,Z$) and in $J$, if the $\gamma$ bands are known to high spin. Unfortunately, in the region around $^{186}$W -- with the heaviest stable isotopes of each element -- the $\gamma$ bands are rarely developed above $J = 6$, as heavy-ion fusion-evaporation reactions cannot be used and only the bandhead region can be explored, usually by Coulomb excitation. 

\begin{figure}[t!]
\begin{center}
   \includegraphics*[trim=0cm 6cm 0cm 0cm, clip=true, width=8.5cm]{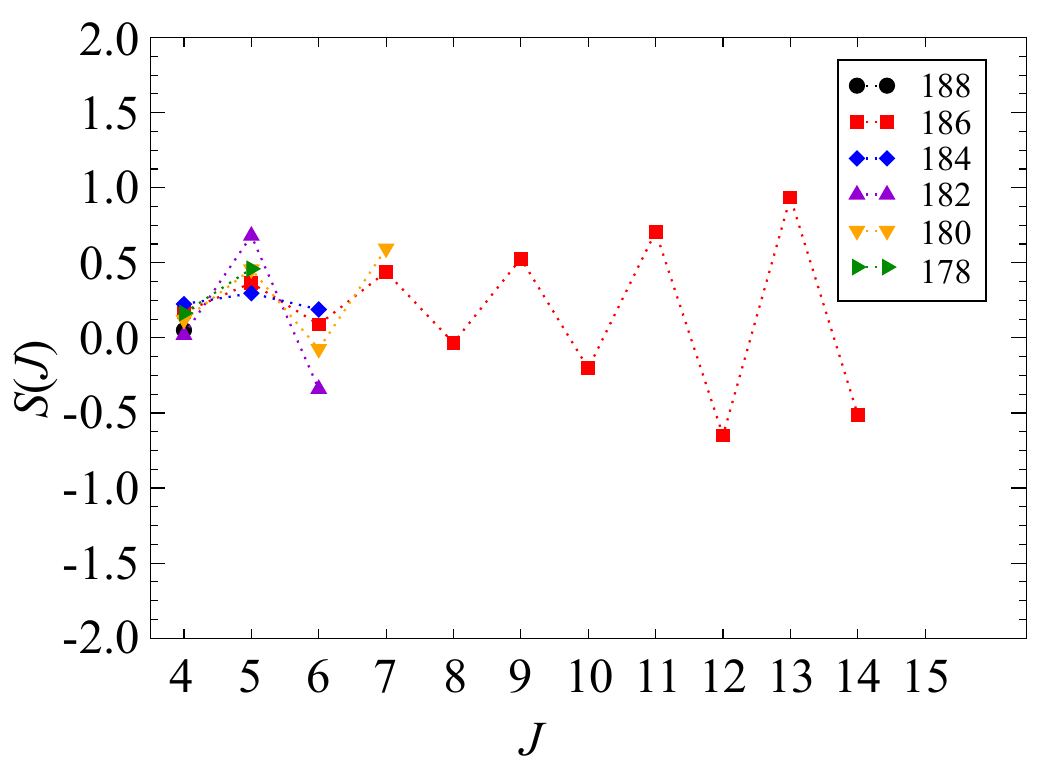} 
\caption{The parameter $S(J)$ plotted as a function of spin for $\gamma$ bands in the heaviest tungsten isotopes; data are from Refs.~\cite{A=178,A=180,A=182,A=184,A=188} (green right-triangles, yellow down-triangles, purple up-triangles, blue diamonds, black circles) and the present study (red squares). The current work allows the spin dependence of triaxiality to be studied in a W isotope for the first time.}
\label{fig12}
\end{center}
\end{figure}

The power of the heavy-ion, inelastic-scattering approach used in the present study is demonstrated in Fig.~\ref{fig12}, where the staggering parameter is displayed for the known $\gamma$ bands in each neutron-rich tungsten isotope. In general, one oscillation of $S(J)$, or less, is observed, since the bands are only known from Coulomb excitation or $\beta$ decay. The heavy-ion collisions of this work, 10-20$\%$ above the Coulomb barrier, delivered increased angular momentum to non-yrast structures. Populating the odd-spin, high-angular-momentum members of the $\gamma$~band in $^{186}$W proved critical for this study. 

\begin{figure}[tb]
\begin{center}
   \includegraphics*[trim=0cm 6cm 0cm 0cm, clip=true, width=8.5cm]{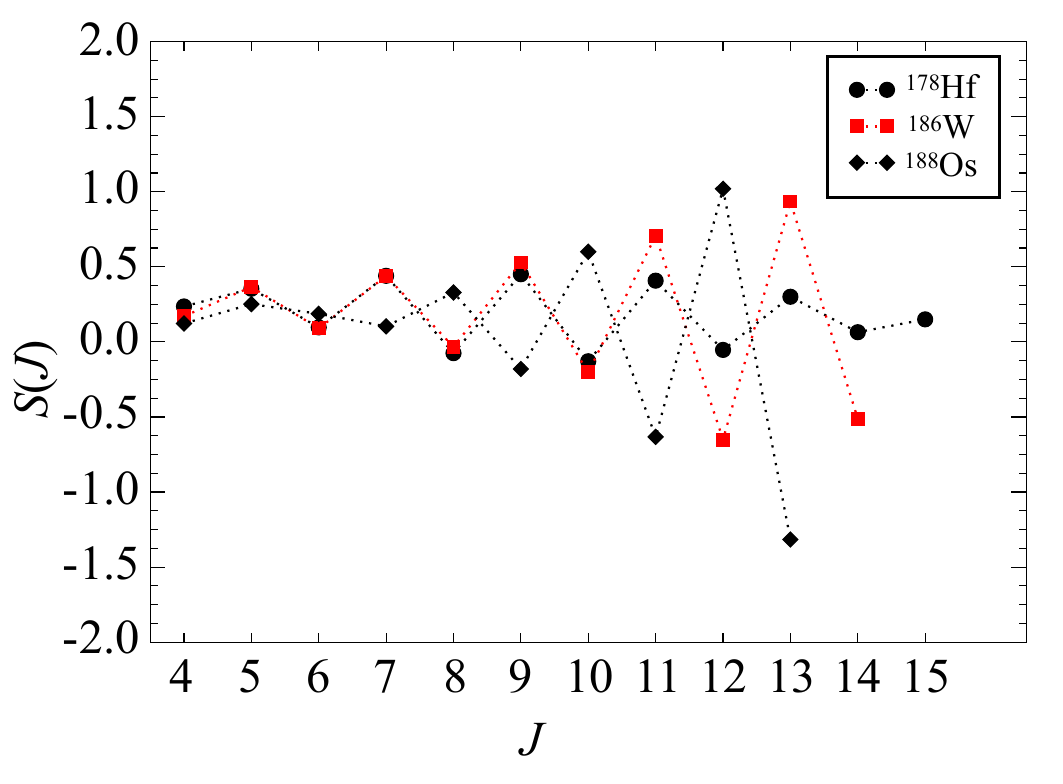} 
\caption{Examples showing the variation of $S(J)$ with spin in $\gamma$ bands of selected stable Hf-W-Os nuclei, including: the staggering in $^{178}$Hf \cite{hayes2002} (black circles), which is similar to the expectation of an axial, $\gamma$-stiff shape; the $^{186}$W data from the present study (red squares), which are consistent with the expected pattern in the case of $\gamma$ softness; and data for $^{188}$Os \cite{modamio2009} (black diamonds), which is proposed to be triaxial at high spin.}
\label{fig13}
\end{center}
\end{figure}

The exceptions in neighboring elements are $^{178,180}$Hf~\cite{hayes2002, tandel2008-2} and $^{186,188}$Os~\cite{wheldon1999, modamio2009}. These cases show distinct patterns that are connected to the underlying physics. Figure~\ref{fig13} illustrates slightly positive, near-constant values of $S(J)$ in $^{178}$Hf, consistent with axial rotation; the current $\gamma$-soft nuclide $^{186}$W, with positive $S(J)$ values for odd-$J$ spins; and (c) $^{188}$Os, which is axial at low spin, with small and slightly positive $S(J)$ values, but transitions to large, positive $S(J)$ values for even spins above $J = 8$, revealing a change to a rigid, triaxial shape. Note that the staggering in $^{188}$Os is out of phase with that seen in $^{178}$Hf and $^{186}$W.

In principle, the even-spin states in the $\gamma$ band can also be perturbed by members of the $K = 0^+$ band. However, in this case -- as can be seen from Fig.~\ref{fig4} -- strong mixing between the $\gamma$ band and ground-state band results in numerous $J \to J$, inter-band transitions, whereas $J~\to~J$ transitions between the $K = 0^+$ band and ground-state band are rather weak, indicating that the mixing is significantly smaller. 

\begin{figure}[t!]
\begin{center}
   \includegraphics*[trim=0cm 6cm 0cm 0cm, clip=true, width=8.5cm]{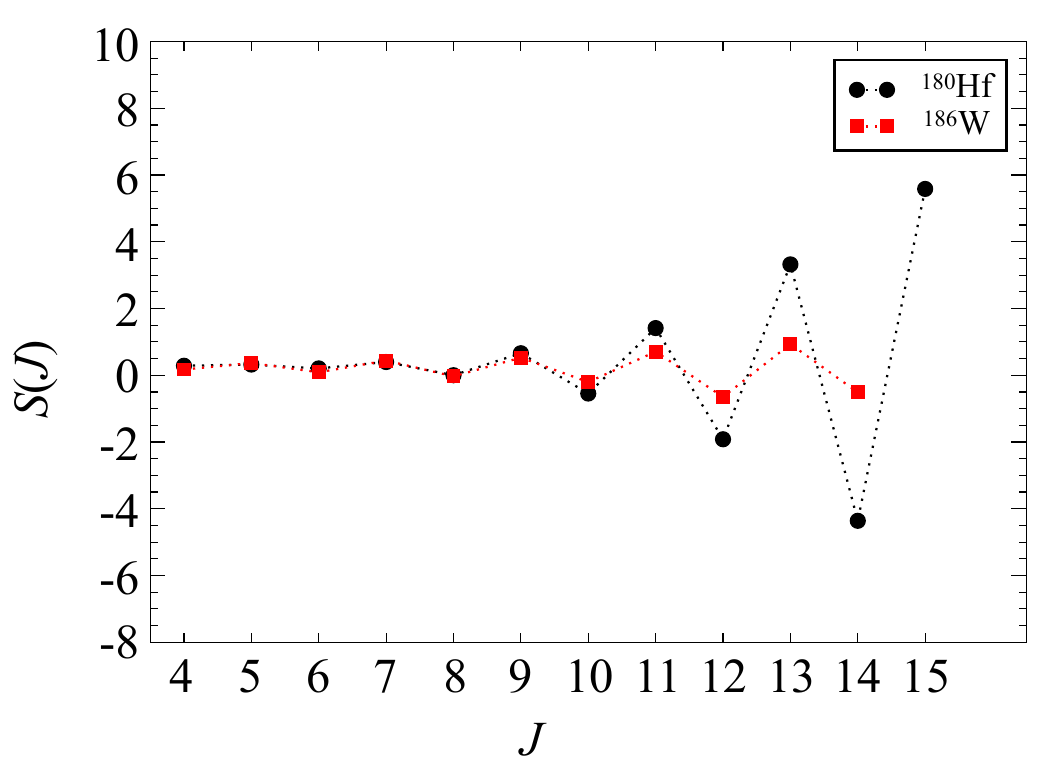} 
\caption{Staggering patterns for the $\gamma$ bands in $^{186}$W from the present study (red squares) and shape co-existence in $^{180}$Hf \cite{A=180, tandel2008-1} (black circles).}
\label{fig14}
\end{center}
\end{figure}

The $S(J)$ parameterization of Refs.~\cite{mccutchan2007,zamfir1991} was developed to enable spectroscopic observables to be directly related to the underlying nuclear potential. In the cases considered in these references, the potential under inspection is assumed to possess a well-defined shape and a single minimum. However, shape co-existence -- with two or more distinct bound shapes -- has been found to be quite ubiquitous across the periodic table \cite{heyde2011}. Interference between the wave functions of states associated with two distinct coexisting shapes may also result in dramatic oscillations of $S(J)$ with spin. This can be seen in the case of $^{180}$Hf \cite{tandel2008-1}, shown here in Fig.~\ref{fig14}. This figure compares the observed $S(J)$ patterns for $^{180}$Hf, where shape coexistence has been proposed, to $^{186}$W from this work. Shape coexistence is predicted to occur in $^{180}$Hf between a near-axially symmetric, prolate-deformed configuration at the ground state and an oblate one located at about 2~MeV in excitation at low spin \cite{hilton1979}. With rotation, the oblate configuration in $^{180}$Hf decreases rapidly in relative excitation energy, crossing the $\gamma$ band at $J \approx 10 \hbar$, before becoming the yrast configuration at $J \approx 20 \hbar$ \cite{tandel2008-1}. Since mixing between two states requires both their spin and parity quantum numbers to be the same, the oblate rotational sequence, with even spin and positive parity, mixes only with the even members of the $\gamma$ band, leaving the odd-spin states unperturbed. Thus, in this interpretation, the states above $10 \hbar$ in this band have changed character from a $\gamma$ vibration to an oblate rotation, and the strong oscillation of $S(J)$ which increases with spin is no longer connected to the physics of rigidity or softness in the $\gamma$ vibrational degree of freedom. Therefore, it is not surprising that this oscillation is much larger in amplitude than any of those calculated with the models reported in Ref.~\cite{mccutchan2007}. 

While a similar interpretation is predicted for $^{186}$W \cite{prasher2015}, given the available statistics and paucity of data up to higher spins, the non-observation of a candidate oblate band at low spins leaves this as a matter of conjecture. It is, however, intriguing to note that, if oblate structures are truly in yrast competition with prolate rotation in these nuclei, this method of populating these structures - inelastic or multi-Coulomb excitation - may contribute to the observed sharp drop in intensities of the ground-state bands in both $^{180}$Hf and $^{186}$W at respective spins of $20 \hbar$ and $14 \hbar$, mirroring the reduced matrix elements connecting states associated with the different shapes. This predicted lowering of the crossover spins at which the oblate configuration becomes yrast was one of the motivations for the present experiment \cite{prasher2015}. Of course, the lowering of the $\gamma$-vibrational bandhead in $^{186}$W by almost a factor of two compared to $^{180}$Hf, indicating significant $\gamma$ softness, could also smear out any clean demarcation of axial prolate and oblate shapes through non-axial degrees of freedom.


\subsection{THE \boldmath{$K^{\pi} = 2^{-}$}, OCTUPOLE BAND}
The Coulomb excitation study in Ref.~\cite{mcgowan1977} used $\alpha$ particles at 25\% below the barrier and was, therefore, dominated by single-step excitations. Levels with large $ E2 $ and $ E3 $ matrix elements linking them to the ground state were strongly populated. In particular, the 1045-keV, $ J^{\pi} = 3^{-}_{\rm{1}} $ state could be studied in detail. An $ E3 $ octupole excitation probability of $ B(E3:0^+_{\rm{1}} \rightarrow 3^-_{\rm{1}}) = 7.0(6) $~W.u. was measured, which represented a considerable fraction of the overall octupole vibrational collectivity \cite{mcgowan1977}. 

In a series of theoretical studies on the coupling of octupole vibrational phonons to deformed nuclei, in both axially symmetric and triaxial systems \cite{neergard1969, neergard1970, vogel1976, toki1977}, it was predicted that, for $^{186}$W, the coupling of the octupole phonon onto the deformation axis with $ K = 2 $ would be lowest in energy, with the $ K = 3 $ coupling located about 40~keV higher, while both the $ K = 1 $ and $ K = 0 $ configurations were calculated to be less bound by several-hundred keV. Considerable Coriolis mixing was also predicted to occur. These findings are broadly supported further by the Interacting Boson Model calculations of Ref.~\cite{cottle1996}.

\begin{figure}[t!]
\begin{center}
   \includegraphics*[trim=0cm 6cm 0cm 0cm, clip=true, width=8.5cm]{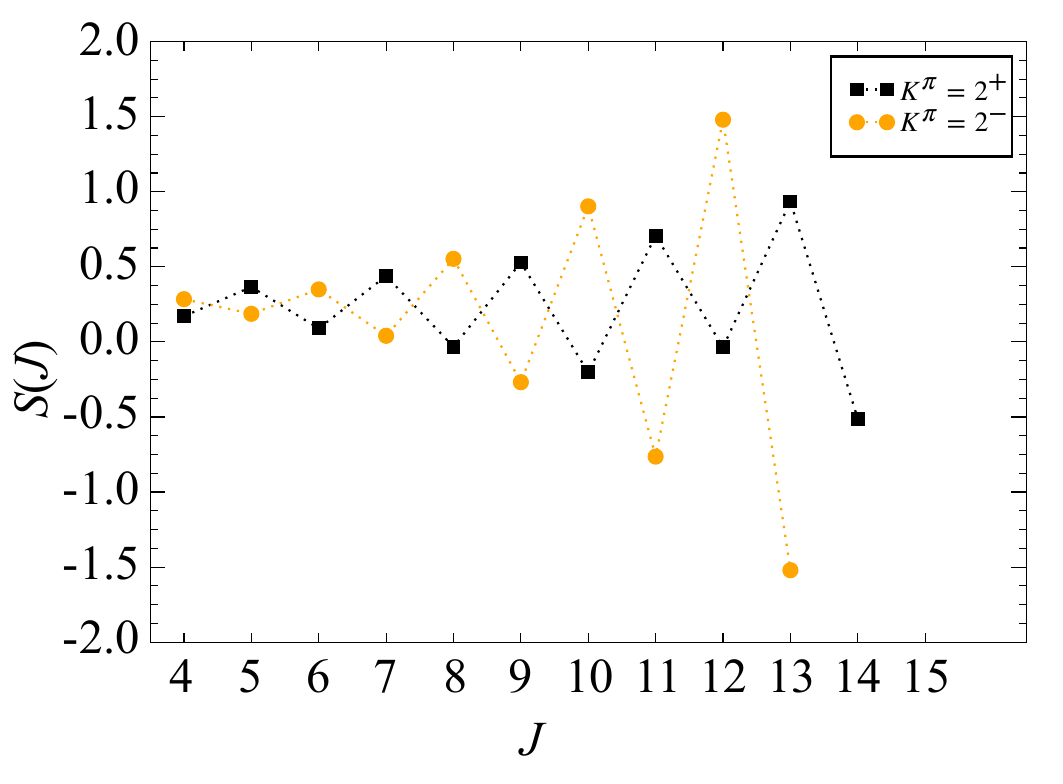} 
\caption{Staggering patterns for the $\gamma$ (black squares) and octupole (orange circles) bands in $^{186}$W, from this work.}
\label{fig15}
\end{center}
\end{figure}

Although the 1045-keV level is directly populated from the ground state by an $ E3 $ transition in Coulomb excitation, it has several faster decay paths to the $ K^{\pi} = 2^{-} $, $ K^{\pi} = 2^{+} $, and $ J^{\pi} = 3^{+} $ bandheads, as well as the $ 2^{+}_{\rm{1}} $ level. An $ E3 $ direct decay back to the ground state was estimated to have a relative branching ratio of approximately 0.01\% of the strongest decay from this $ 3^{-}_{\rm{1}} $ level, a value far below the experimental sensitivity.

Alaga $et~al$. \cite{alaga1955} pointed out that the ratio of decay matrix elements from deformed states such as these should depend solely on angular-momentum geometry, e.g. on the changes in the directions in which the angular momentum vectors point relative to the deformation axis. As such, they are sensitive to the angular momentum projections, $ K $. These ratios can be reduced to a ratio of Clebsch-Gordan coefficients for decays from any state to members of another rotational sequence \cite{mcgowan1977}. A test of this hypothesis was made in Ref.~\cite{mcgowan1977} by measuring the electric dipole branches from the $ J^{\pi} = 3^{-}_{\rm{1}} $ state at 1045~keV, with the specific goal of addressing whether this level is the $ J = 3 $ member of a strongly coupled $ K = 2 $ band, or the bandhead of the expected $ J = K = 3 $ band. The experimental probe was to measure the decays to the well-known, $ K^{\pi} = 2^{+} $ bandhead, and use the ratio of reduced transition probabilities, $R$, such that according to the so-called Alaga rules \cite{alaga1955}:\\[-0.6cm]
 
\begin{equation}
R = \frac{B(E1: 3^-_{\rm{1}} \to 2^+_{\rm{2}})}{B(E1:3^-_{\rm{1}} \to 3^+_{\rm{1}})}=\frac{<3 1 K_i (2-K_i) | 3 1 2 2 >^{2}}{< 3 1 K_i (2-K_i) | 3 1 3 2 >^{2}}.
\label{equ:mratio}
\end{equation}

\noindent 
The reduced transition probabilities, $ B(EL) $, are proportional to the partial $\gamma$-ray branches, $ Br_{\gamma} $, from the decaying level through the relation: \\[-0.6cm]

\begin{equation}
R = \frac{ Br_{\gamma}(3^-_{\rm{1}} \to 2^+_{\rm{2}}) \times E_{\gamma}(3^-_{\rm{1}} \to 3^+_{\rm{1}})^3 }{Br_{\gamma}(3^-_{\rm{1}} \to 3^+_{\rm{1}})\times E_{\gamma}(3^-_{\rm{1}} \to 2^+_{\rm{2}})^3}. \\
\label{equ:conv}
\end{equation}

\begin{figure}[t!]
\begin{center}
   \includegraphics*[trim=0cm 6cm 0cm 0cm, clip=true, width=8.5cm]{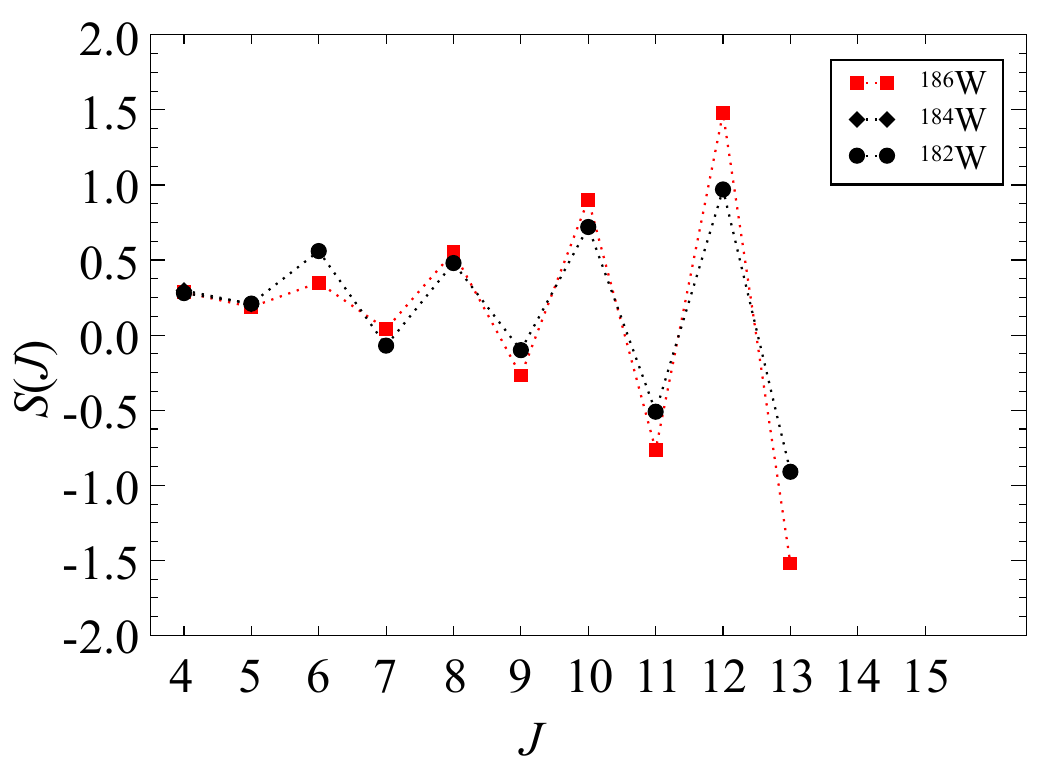} 
\caption{Staggering patterns for the octupole bands in $^{186}$W from this work (red squares), $^{184}$W \cite{A=184} (black diamonds), and $^{182}$W \cite{A=182} (black circles), illustrating their striking similarity.}
\label{fig16}
\end{center}
\end{figure}

\noindent
If $ K_i = 2 $, this expression gives $ R = 0.71 $, while it is $ R = 2.86 $ for $ K_i = 3 $. The experimentally measured value of 0.69(3) \cite{mcgowan1977} showed agreement with a $ K = 2 $ assignment for the $ J^{\pi} = 3^{-}_{\rm{1}} $ state, despite the fact that the calculations in Ref.~\cite{neergard1970} suggest almost complete mixing between the $ K = 2 $ and 3 states. With the new data from this work, the measurement of Ref.~\cite{mcgowan1977} can be reassessed and more extensive tests of the Alaga rules on the decays of other negative-parity states can be performed to ascertain whether all the states are consistent with the pure $ K = 2 $ assignment. This is particularly important for the odd-spin sequence, and especially so if the ratios are found to deviate from the Alaga-rule predictions with increasing spin. \\

\textit{The $J^{\pi} = 3^{-}_{\rm{1}} $ state}: This level decays to the $\gamma$ band via $ 3^{-}_{\rm{1}} \to 3^{+}_{\rm{1}} $, and $ 3^{-}_{\rm{1}} \to 2^{+}_{\rm{2}} $, $ E1 $ branches by $\gamma$ rays of energies 183 and 308~keV, respectively. Using the intensities given in Table~\ref{table1}, a ratio $ R = 0.66(2) $ is calculated. This is consistent with the prior value and it is 90\% of the Alaga prediction of 0.71 for $ K = 2 $; axial asymmetry and softness described above could explain the attenuation.

\textit{The $J^{\pi} = 4^{-}_{\rm{1}} $ state}: This state decays to the $\gamma$ band via $ 4^{-}_{\rm{1}} \to 4^{+}_{\rm{2}} $, and $ 4^{-}_{\rm{1}} \to 3^{+}_{\rm{1}} $, $ E1 $ branches of 165 and 309~keV, respectively, and a value $ R = 1.50(8) $ is measured. The Alaga prediction is $ R = 1.66 $ for both the $ K = 2 $ and $ K = 3 $ cases. Therefore, while this result provides a consistency test of the Alaga rule, again $\approx$ 90\% of the prediction, it does not constrain the value of $ K $. 

\textit{The $J^{\pi} = 5^{-}_{\rm{1}} $ state}: This level is expected to decay to the $ \gamma $ band via $ 5^{-}_{\rm{1}} \to 5^{+}_{\rm{1}} $ and $ 5^{-}_{\rm{1}} \to 4^{+}_{\rm{2}} $ transitions of 125 and 315~keV, respectively. The $ 5^{-}_{\rm{1}} \to 5^{+}_{\rm{1}} $ transition energy is close to that of the very intense, first excited-state decay in the ground-state band at 123~keV, and clear proof of its existence could not be isolated. The Alaga prediction for the branching ratio is again sensitive to $ K $, and has values of $ R = 2.86 $ for $ K = 2 $ and $ R = 1.28 $ for $ K = 3 $, respectively. Even in the $K = 2$ case, where the branch would be strongest, the expected $\gamma$-ray intensity is $< 1.5$ units (see Tab.~\ref{table1}), which is more than a factor of two below the experimental sensitivity. 

Overall, the measured $\gamma$-decay branching ratios are consistent with the notion that bands 4 and 5 are the signature partners of a strongly coupled, $ K^{\pi} = 2^{-} $ octupole vibrational band. The noted reduction below the Alaga prediction is consistent with a non-axial potential. More significantly, there is again a clear odd-spin, even-spin staggering of the level energies. To illustrate this perturbation, the staggering parameter, $ S(J) $ from Eq.~\ref{equ:stag}, is again useful. Figure~\ref{fig15} shows the staggering pattern of the octupole band of $^{186}$W; it is out of phase with the pattern found in the $\gamma$ band and has a different underlying cause. In this case, the perturbing interaction is through Coriolis-driven mixing between different projections of the octupole vibrational phonon. This effect was explored in detail in $^{176}$Hf \cite{khoo1976}; it was found to mainly involve mixing with the higher-lying $K = 0^-$ band, which only has levels with odd spins. These odd-spin, negative-parity states mix with the odd-spin members of the $ K^{\pi} = 2^{-} $ band and depress them. The spin dependence of this interaction is proportional to the classical Coriolis interaction, which increases with spin in a smooth and predictable fashion. 

The staggering pattern observed in strongly coupled octupole bands appears to be quite ubiquitous, and can be seen in the even-even tungsten isotopes where data are available, as highlighted in Fig.~\ref{fig16}. Several similar cases have also recently been found in axially deformed actinide nuclei \cite{qiu2016}. To investigate this Coriolis coupling in $^{186}$W in a more quantitative way, the locations of the other bandheads are needed, particularly the excitation energy of the $ K^{\pi} = 3^- $ band, and the predicted \cite{neergard1970}, but not yet observed, $ K^{\pi} = 0^- $ band. With contemporary detector set-ups this should be possible using light-ion inelastic scattering. \\



\section{SUMMARY AND CONCLUSIONS}
Inelastic scattering of $^{136}$Xe at 725 and 800 MeV (10 and 20\% above the Coulomb barrier) was found to populate non-yrast states in $^{186}$W to intermediate spin. In particular, the odd-spin members of the $\gamma$ band were extended to $ J^{\pi} = 11^+_{\rm{1}} $, and the new data provide insight into triaxial softness and its evolution with spin in this nucleus. The odd-even staggering pattern in the $\gamma$ band was found to be consistent with a potential that gets softer in the $\gamma$ degree of freedom with increasing spin. The signature partners of the strongly coupled, $ K^{\pi} = 2^{-} $, octupole band were extended to $ J^{\pi} = 12^-_{\rm{1}} $ and $ J^{\pi} = 13^-_{\rm{1}} $, respectively. This band was also found to have odd-even staggering, but with a phase opposite to that of the $\gamma$ band. This staggering is associated with Coriolis coupling with other, unobserved, octupole bands. The present results provide another example of heavy-ion, inelastic-scattering reactions being a powerful tool for non-yrast, nuclear-structure physics when more traditional fusion-evaporation reactions are unavailable. This approach should also work well for experiments with re-accelerated radioactive beams. \\



\section*{ACKNOWLEDGMENTS}

The authors wish to acknowledge the excellent work of the Physics Support group of the ATLAS Facility at Argonne National Laboratory. This material is based upon work supported by the U.S. Department of Energy, Office of Science, Office of Nuclear Physics under Grants No.~DE-FG02-94ER40848 (UML), No.~DE-FG02-97ER41041 (UNC), No. DE-FG02-97ER41033 (TUNL) and DE-FG02-94-ER40834 (UMCP), and Contracts No.~DE-AC02-06CH11357 (ANL) and No.~DE-AC52-07NA27344 (LLNL), the International Technology Center Pacific (ITC-PAC) under Contract No.~FA520919PA138 (ANU), and the National Science Foundation. The research used resources of ANL's ATLAS facility, which is a DOE Office of Science user facility.\\





\begin{thebibliography}{56}%
\makeatletter
\providecommand \@ifxundefined [1]{%
 \@ifx{#1\undefined}
}%
\providecommand \@ifnum [1]{%
 \ifnum #1\expandafter \@firstoftwo
 \else \expandafter \@secondoftwo
 \fi
}%
\providecommand \@ifx [1]{%
 \ifx #1\expandafter \@firstoftwo
 \else \expandafter \@secondoftwo
 \fi
}%
\providecommand \natexlab [1]{#1}%
\providecommand \enquote  [1]{``#1''}%
\providecommand \bibnamefont  [1]{#1}%
\providecommand \bibfnamefont [1]{#1}%
\providecommand \citenamefont [1]{#1}%
\providecommand \href@noop [0]{\@secondoftwo}%
\providecommand \href [0]{\begingroup \@sanitize@url \@href}%
\providecommand \@href[1]{\@@startlink{#1}\@@href}%
\providecommand \@@href[1]{\endgroup#1\@@endlink}%
\providecommand \@sanitize@url [0]{\catcode `\\12\catcode `\$12\catcode
  `\&12\catcode `\#12\catcode `\^12\catcode `\_12\catcode `\%12\relax}%
\providecommand \@@startlink[1]{}%
\providecommand \@@endlink[0]{}%
\providecommand \url  [0]{\begingroup\@sanitize@url \@url }%
\providecommand \@url [1]{\endgroup\@href {#1}{\urlprefix }}%
\providecommand \urlprefix  [0]{URL }%
\providecommand \Eprint [0]{\href }%
\providecommand \doibase [0]{http://dx.doi.org/}%
\providecommand \selectlanguage [0]{\@gobble}%
\providecommand \bibinfo  [0]{\@secondoftwo}%
\providecommand \bibfield  [0]{\@secondoftwo}%
\providecommand \translation [1]{[#1]}%
\providecommand \BibitemOpen [0]{}%
\providecommand \bibitemStop [0]{}%
\providecommand \bibitemNoStop [0]{.\EOS\space}%
\providecommand \EOS [0]{\spacefactor3000\relax}%
\providecommand \BibitemShut  [1]{\csname bibitem#1\endcsname}%
\let\auto@bib@innerbib\@empty
\bibitem [{\citenamefont {Regan}\ \emph {et~al.}(2002)\citenamefont {Regan},
  \citenamefont {Xu}, \citenamefont {Walker}, \citenamefont {Oi}, \citenamefont
  {Rath},\ and\ \citenamefont {Stevenson}}]{regan2002}%
  \BibitemOpen
  \bibfield  {author} {\bibinfo {author} {\bibfnamefont {P.~H.}\ \bibnamefont
  {Regan}}, \bibinfo {author} {\bibfnamefont {F.~R.}\ \bibnamefont {Xu}},
  \bibinfo {author} {\bibfnamefont {P.~M.}\ \bibnamefont {Walker}}, \bibinfo
  {author} {\bibfnamefont {M.}~\bibnamefont {Oi}}, \bibinfo {author}
  {\bibfnamefont {A.~K.}\ \bibnamefont {Rath}}, \ and\ \bibinfo {author}
  {\bibfnamefont {P.~D.}\ \bibnamefont {Stevenson}},\ }\href {\doibase
  10.1103/PhysRevC.65.037302} {\bibfield  {journal} {\bibinfo  {journal} {Phys.
  Rev. C}\ }\textbf {\bibinfo {volume} {65}},\ \bibinfo {pages} {037302}
  (\bibinfo {year} {2002})}\BibitemShut {NoStop}%
\bibitem [{\citenamefont {Heusler}\ \emph {et~al.}(2016)\citenamefont
  {Heusler}, \citenamefont {Jolos}, \citenamefont {Faestermann}, \citenamefont
  {Hertenberger}, \citenamefont {Wirth},\ and\ \citenamefont {von
  Brentano}}]{heusler2016}%
  \BibitemOpen
  \bibfield  {author} {\bibinfo {author} {\bibfnamefont {A.}~\bibnamefont
  {Heusler}}, \bibinfo {author} {\bibfnamefont {R.~V.}\ \bibnamefont {Jolos}},
  \bibinfo {author} {\bibfnamefont {T.}~\bibnamefont {Faestermann}}, \bibinfo
  {author} {\bibfnamefont {R.}~\bibnamefont {Hertenberger}}, \bibinfo {author}
  {\bibfnamefont {H.-F.}\ \bibnamefont {Wirth}}, \ and\ \bibinfo {author}
  {\bibfnamefont {P.}~\bibnamefont {von Brentano}},\ }\href {\doibase
  10.1103/PhysRevC.93.054321} {\bibfield  {journal} {\bibinfo  {journal} {Phys.
  Rev. C}\ }\textbf {\bibinfo {volume} {93}},\ \bibinfo {pages} {054321}
  (\bibinfo {year} {2016})}\BibitemShut {NoStop}%
\bibitem [{\citenamefont {Kumar}\ and\ \citenamefont
  {Baranger}(1968)}]{kumar1968}%
  \BibitemOpen
  \bibfield  {author} {\bibinfo {author} {\bibfnamefont {K.}~\bibnamefont
  {Kumar}}\ and\ \bibinfo {author} {\bibfnamefont {M.}~\bibnamefont
  {Baranger}},\ }\href {\doibase https://doi.org/10.1016/0375-9474(68)90371-0}
  {\bibfield  {journal} {\bibinfo  {journal} {Nucl. Phys. A}\ }\textbf
  {\bibinfo {volume} {110}},\ \bibinfo {pages} {529} (\bibinfo {year}
  {1968})}\BibitemShut {NoStop}%
\bibitem [{\citenamefont {Sarriguren}\ \emph {et~al.}(2008)\citenamefont
  {Sarriguren}, \citenamefont {Rodr\'{\i}guez-Guzm\'an},\ and\ \citenamefont
  {Robledo}}]{sariguren2008}%
  \BibitemOpen
  \bibfield  {author} {\bibinfo {author} {\bibfnamefont {P.}~\bibnamefont
  {Sarriguren}}, \bibinfo {author} {\bibfnamefont {R.}~\bibnamefont
  {Rodr\'{\i}guez-Guzm\'an}}, \ and\ \bibinfo {author} {\bibfnamefont {L.~M.}\
  \bibnamefont {Robledo}},\ }\href {\doibase 10.1103/PhysRevC.77.064322}
  {\bibfield  {journal} {\bibinfo  {journal} {Phys. Rev. C}\ }\textbf {\bibinfo
  {volume} {77}},\ \bibinfo {pages} {064322} (\bibinfo {year}
  {2008})}\BibitemShut {NoStop}%
\bibitem [{\citenamefont {Stevenson}\ \emph {et~al.}(2005)\citenamefont
  {Stevenson}, \citenamefont {Brine}, \citenamefont {Podolyak}, \citenamefont
  {Regan}, \citenamefont {Walker},\ and\ \citenamefont
  {Stone}}]{stevenson2005}%
  \BibitemOpen
  \bibfield  {author} {\bibinfo {author} {\bibfnamefont {P.~D.}\ \bibnamefont
  {Stevenson}}, \bibinfo {author} {\bibfnamefont {M.~P.}\ \bibnamefont
  {Brine}}, \bibinfo {author} {\bibfnamefont {Z.}~\bibnamefont {Podolyak}},
  \bibinfo {author} {\bibfnamefont {P.~H.}\ \bibnamefont {Regan}}, \bibinfo
  {author} {\bibfnamefont {P.~M.}\ \bibnamefont {Walker}}, \ and\ \bibinfo
  {author} {\bibfnamefont {J.~R.}\ \bibnamefont {Stone}},\ }\href {\doibase
  10.1103/PhysRevC.72.047303} {\bibfield  {journal} {\bibinfo  {journal} {Phys.
  Rev. C}\ }\textbf {\bibinfo {volume} {72}},\ \bibinfo {pages} {047303}
  (\bibinfo {year} {2005})}\BibitemShut {NoStop}%
\bibitem [{\citenamefont {Casten}\ \emph {et~al.}(1973)\citenamefont {Casten},
  \citenamefont {Keaton},\ and\ \citenamefont {Lawrence}}]{casten1973}%
  \BibitemOpen
  \bibfield  {author} {\bibinfo {author} {\bibfnamefont {R.~F.}\ \bibnamefont
  {Casten}}, \bibinfo {author} {\bibfnamefont {P.~W.}\ \bibnamefont {Keaton}},
  \ and\ \bibinfo {author} {\bibfnamefont {G.~P.}\ \bibnamefont {Lawrence}},\
  }\href {\doibase 10.1103/PhysRevC.7.1016} {\bibfield  {journal} {\bibinfo
  {journal} {Phys. Rev. C}\ }\textbf {\bibinfo {volume} {7}},\ \bibinfo {pages}
  {1016} (\bibinfo {year} {1973})}\BibitemShut {NoStop}%
\bibitem [{\citenamefont {Basunia}(2006)}]{A=176}%
  \BibitemOpen
  \bibfield  {author} {\bibinfo {author} {\bibfnamefont {M.~S.}\ \bibnamefont
  {Basunia}},\ }\href {\doibase https://doi.org/10.1016/j.nds.2006.03.001}
  {\bibfield  {journal} {\bibinfo  {journal} {Nucl. Data Sheets}\ }\textbf
  {\bibinfo {volume} {107}},\ \bibinfo {pages} {791} (\bibinfo {year}
  {2006})}\BibitemShut {NoStop}%
\bibitem [{\citenamefont {Achterberg}\ \emph {et~al.}(2009)\citenamefont
  {Achterberg}, \citenamefont {Capurro},\ and\ \citenamefont {Marti}}]{A=178}%
  \BibitemOpen
  \bibfield  {author} {\bibinfo {author} {\bibfnamefont {E.}~\bibnamefont
  {Achterberg}}, \bibinfo {author} {\bibfnamefont {O.~A.}\ \bibnamefont
  {Capurro}}, \ and\ \bibinfo {author} {\bibfnamefont {G.~V.}\ \bibnamefont
  {Marti}},\ }\href {\doibase https://doi.org/10.1016/j.nds.2009.05.002}
  {\bibfield  {journal} {\bibinfo  {journal} {Nucl. Data Sheets}\ }\textbf
  {\bibinfo {volume} {110}},\ \bibinfo {pages} {1473} (\bibinfo {year}
  {2009})}\BibitemShut {NoStop}%
\bibitem [{\citenamefont {McCutchan}(2015)}]{A=180}%
  \BibitemOpen
  \bibfield  {author} {\bibinfo {author} {\bibfnamefont {E.~A.}\ \bibnamefont
  {McCutchan}},\ }\href {\doibase https://doi.org/10.1016/j.nds.2015.05.002}
  {\bibfield  {journal} {\bibinfo  {journal} {Nucl. Data Sheets}\ }\textbf
  {\bibinfo {volume} {126}},\ \bibinfo {pages} {151} (\bibinfo {year}
  {2015})}\BibitemShut {NoStop}%
\bibitem [{\citenamefont {Singh}(2015)}]{A=182}%
  \BibitemOpen
  \bibfield  {author} {\bibinfo {author} {\bibfnamefont {B.}~\bibnamefont
  {Singh}},\ }\href {\doibase https://doi.org/10.1016/j.nds.2015.11.002}
  {\bibfield  {journal} {\bibinfo  {journal} {Nucl. Data Sheets}\ }\textbf
  {\bibinfo {volume} {130}},\ \bibinfo {pages} {21} (\bibinfo {year}
  {2015})}\BibitemShut {NoStop}%
\bibitem [{\citenamefont {Baglin}(2010)}]{A=184}%
  \BibitemOpen
  \bibfield  {author} {\bibinfo {author} {\bibfnamefont {C.~M.}\ \bibnamefont
  {Baglin}},\ }\href {\doibase https://doi.org/10.1016/j.nds.2010.01.001}
  {\bibfield  {journal} {\bibinfo  {journal} {Nucl. Data Sheets}\ }\textbf
  {\bibinfo {volume} {111}},\ \bibinfo {pages} {275} (\bibinfo {year}
  {2010})}\BibitemShut {NoStop}%
\bibitem [{\citenamefont {Baglin}(2003)}]{A=186}%
  \BibitemOpen
  \bibfield  {author} {\bibinfo {author} {\bibfnamefont {C.~M.}\ \bibnamefont
  {Baglin}},\ }\href {\doibase https://doi.org/10.1006/ndsh.2003.0007}
  {\bibfield  {journal} {\bibinfo  {journal} {Nucl. Data Sheets}\ }\textbf
  {\bibinfo {volume} {99}},\ \bibinfo {pages} {1} (\bibinfo {year}
  {2003})}\BibitemShut {NoStop}%
\bibitem [{\citenamefont {Kondev}\ \emph {et~al.}(2018)\citenamefont {Kondev},
  \citenamefont {Juutinen},\ and\ \citenamefont {Hartley}}]{A=188}%
  \BibitemOpen
  \bibfield  {author} {\bibinfo {author} {\bibfnamefont {F.~G.}\ \bibnamefont
  {Kondev}}, \bibinfo {author} {\bibfnamefont {S.}~\bibnamefont {Juutinen}}, \
  and\ \bibinfo {author} {\bibfnamefont {D.~J.}\ \bibnamefont {Hartley}},\
  }\href {\doibase https://doi.org/10.1016/j.nds.2018.05.001} {\bibfield
  {journal} {\bibinfo  {journal} {Nucl. Data Sheets}\ }\textbf {\bibinfo
  {volume} {150}},\ \bibinfo {pages} {1} (\bibinfo {year} {2018})}\BibitemShut
  {NoStop}%
\bibitem [{\citenamefont {Dracoulis}\ \emph {et~al.}(2016)\citenamefont
  {Dracoulis}, \citenamefont {Walker},\ and\ \citenamefont
  {Kondev}}]{dracoulis2016}%
  \BibitemOpen
  \bibfield  {author} {\bibinfo {author} {\bibfnamefont {G.~D.}\ \bibnamefont
  {Dracoulis}}, \bibinfo {author} {\bibfnamefont {P.~M.}\ \bibnamefont
  {Walker}}, \ and\ \bibinfo {author} {\bibfnamefont {F.~G.}\ \bibnamefont
  {Kondev}},\ }\href {\doibase 10.1088/0034-4885/79/7/076301} {\bibfield
  {journal} {\bibinfo  {journal} {Rep. Prog. Phys.}\ }\textbf {\bibinfo
  {volume} {79}},\ \bibinfo {pages} {076301} (\bibinfo {year}
  {2016})}\BibitemShut {NoStop}%
\bibitem [{\citenamefont {Purry}\ \emph {et~al.}(1998)\citenamefont {Purry},
  \citenamefont {Walker}, \citenamefont {Dracoulis}, \citenamefont
  {Kib\'{e}di}, \citenamefont {Kondev}, \citenamefont {Bayer}, \citenamefont
  {Bruce}, \citenamefont {Byrne}, \citenamefont {Gelletly}, \citenamefont
  {Regan}, \citenamefont {Thwaites}, \citenamefont {Burglin},\ and\
  \citenamefont {Rowley}}]{purry1998}%
  \BibitemOpen
  \bibfield  {author} {\bibinfo {author} {\bibfnamefont {C.~S.}\ \bibnamefont
  {Purry}}, \bibinfo {author} {\bibfnamefont {P.~M.}\ \bibnamefont {Walker}},
  \bibinfo {author} {\bibfnamefont {G.~D.}\ \bibnamefont {Dracoulis}}, \bibinfo
  {author} {\bibfnamefont {T.}~\bibnamefont {Kib\'{e}di}}, \bibinfo {author}
  {\bibfnamefont {F.~G.}\ \bibnamefont {Kondev}}, \bibinfo {author}
  {\bibfnamefont {S.}~\bibnamefont {Bayer}}, \bibinfo {author} {\bibfnamefont
  {A.~M.}\ \bibnamefont {Bruce}}, \bibinfo {author} {\bibfnamefont {A.~P.}\
  \bibnamefont {Byrne}}, \bibinfo {author} {\bibfnamefont {W.}~\bibnamefont
  {Gelletly}}, \bibinfo {author} {\bibfnamefont {P.~H.}\ \bibnamefont {Regan}},
  \bibinfo {author} {\bibfnamefont {C.}~\bibnamefont {Thwaites}}, \bibinfo
  {author} {\bibfnamefont {O.}~\bibnamefont {Burglin}}, \ and\ \bibinfo
  {author} {\bibfnamefont {N.}~\bibnamefont {Rowley}},\ }\href {\doibase
  https://doi.org/10.1016/S0375-9474(97)00654-4} {\bibfield  {journal}
  {\bibinfo  {journal} {Nucl. Phys. A}\ }\textbf {\bibinfo {volume} {632}},\
  \bibinfo {pages} {229} (\bibinfo {year} {1998})}\BibitemShut {NoStop}%
\bibitem [{\citenamefont {Wheldon}\ \emph {et~al.}(1998)\citenamefont
  {Wheldon}, \citenamefont {D'Alarcao}, \citenamefont {Chowdhury},
  \citenamefont {Walker}, \citenamefont {Seabury}, \citenamefont {Ahmad},
  \citenamefont {Carpenter}, \citenamefont {Cullen}, \citenamefont {Hackman},
  \citenamefont {Janssens}, \citenamefont {Khoo}, \citenamefont {Nisius},
  \citenamefont {Pearson},\ and\ \citenamefont {Reiter}}]{wheldon1998}%
  \BibitemOpen
  \bibfield  {author} {\bibinfo {author} {\bibfnamefont {C.}~\bibnamefont
  {Wheldon}}, \bibinfo {author} {\bibfnamefont {R.}~\bibnamefont {D'Alarcao}},
  \bibinfo {author} {\bibfnamefont {P.}~\bibnamefont {Chowdhury}}, \bibinfo
  {author} {\bibfnamefont {P.~M.}\ \bibnamefont {Walker}}, \bibinfo {author}
  {\bibfnamefont {E.}~\bibnamefont {Seabury}}, \bibinfo {author} {\bibfnamefont
  {I.}~\bibnamefont {Ahmad}}, \bibinfo {author} {\bibfnamefont {M.~P.}\
  \bibnamefont {Carpenter}}, \bibinfo {author} {\bibfnamefont {D.~M.}\
  \bibnamefont {Cullen}}, \bibinfo {author} {\bibfnamefont {G.}~\bibnamefont
  {Hackman}}, \bibinfo {author} {\bibfnamefont {R.~V.~F.}\ \bibnamefont
  {Janssens}}, \bibinfo {author} {\bibfnamefont {T.~L.}\ \bibnamefont {Khoo}},
  \bibinfo {author} {\bibfnamefont {D.}~\bibnamefont {Nisius}}, \bibinfo
  {author} {\bibfnamefont {C.~J.}\ \bibnamefont {Pearson}}, \ and\ \bibinfo
  {author} {\bibfnamefont {P.}~\bibnamefont {Reiter}},\ }\href {\doibase
  https://doi.org/10.1016/S0370-2693(98)00301-3} {\bibfield  {journal}
  {\bibinfo  {journal} {Phys. Lett. B}\ }\textbf {\bibinfo {volume} {425}},\
  \bibinfo {pages} {239} (\bibinfo {year} {1998})}\BibitemShut {NoStop}%
\bibitem [{\citenamefont {Lane}\ \emph {et~al.}(2010)\citenamefont {Lane},
  \citenamefont {Dracoulis}, \citenamefont {Kondev}, \citenamefont {Hughes},
  \citenamefont {Watanabe}, \citenamefont {Byrne}, \citenamefont {Carpenter},
  \citenamefont {Chiara}, \citenamefont {Chowdhury}, \citenamefont {Janssens},
  \citenamefont {Lauritsen}, \citenamefont {Lister}, \citenamefont {McCutchan},
  \citenamefont {Seweryniak}, \citenamefont {Stefanescu},\ and\ \citenamefont
  {Zhu}}]{lane2010}%
  \BibitemOpen
  \bibfield  {author} {\bibinfo {author} {\bibfnamefont {G.~J.}\ \bibnamefont
  {Lane}}, \bibinfo {author} {\bibfnamefont {G.~D.}\ \bibnamefont {Dracoulis}},
  \bibinfo {author} {\bibfnamefont {F.~G.}\ \bibnamefont {Kondev}}, \bibinfo
  {author} {\bibfnamefont {R.~O.}\ \bibnamefont {Hughes}}, \bibinfo {author}
  {\bibfnamefont {H.}~\bibnamefont {Watanabe}}, \bibinfo {author}
  {\bibfnamefont {A.~P.}\ \bibnamefont {Byrne}}, \bibinfo {author}
  {\bibfnamefont {M.~P.}\ \bibnamefont {Carpenter}}, \bibinfo {author}
  {\bibfnamefont {C.~J.}\ \bibnamefont {Chiara}}, \bibinfo {author}
  {\bibfnamefont {P.}~\bibnamefont {Chowdhury}}, \bibinfo {author}
  {\bibfnamefont {R.~V.~F.}\ \bibnamefont {Janssens}}, \bibinfo {author}
  {\bibfnamefont {T.}~\bibnamefont {Lauritsen}}, \bibinfo {author}
  {\bibfnamefont {C.~J.}\ \bibnamefont {Lister}}, \bibinfo {author}
  {\bibfnamefont {E.~A.}\ \bibnamefont {McCutchan}}, \bibinfo {author}
  {\bibfnamefont {D.}~\bibnamefont {Seweryniak}}, \bibinfo {author}
  {\bibfnamefont {I.}~\bibnamefont {Stefanescu}}, \ and\ \bibinfo {author}
  {\bibfnamefont {S.}~\bibnamefont {Zhu}},\ }\href {\doibase
  10.1103/PhysRevC.82.051304} {\bibfield  {journal} {\bibinfo  {journal} {Phys.
  Rev. C}\ }\textbf {\bibinfo {volume} {82}},\ \bibinfo {pages} {051304(R)}
  (\bibinfo {year} {2010})}\BibitemShut {NoStop}%
\bibitem [{\citenamefont {Alkhomashi}\ \emph {et~al.}(2009)\citenamefont
  {Alkhomashi}, \citenamefont {Regan}, \citenamefont {Podoly\'ak},
  \citenamefont {Pietri}, \citenamefont {Garnsworthy}, \citenamefont {Steer},
  \citenamefont {Benlliure}, \citenamefont {Caserejos}, \citenamefont {Casten},
  \citenamefont {Gerl}, \citenamefont {Wollersheim}, \citenamefont {Grebosz},
  \citenamefont {Farrelly}, \citenamefont {G\'orska}, \citenamefont
  {Kojouharov}, \citenamefont {Schaffner}, \citenamefont {Algora},
  \citenamefont {Benzoni}, \citenamefont {Blazhev}, \citenamefont {Boutachkov},
  \citenamefont {Bruce}, \citenamefont {Denis~Bacelar}, \citenamefont {Cullen},
  \citenamefont {C\'aceres}, \citenamefont {Doornenbal}, \citenamefont
  {Estevez}, \citenamefont {Fujita}, \citenamefont {Gelletly}, \citenamefont
  {Hoischen}, \citenamefont {Kumar}, \citenamefont {Kurz}, \citenamefont
  {Lalkovski}, \citenamefont {Liu}, \citenamefont {Mihai}, \citenamefont
  {Molina}, \citenamefont {Morales}, \citenamefont {M\"ucher}, \citenamefont
  {Prokopowicz}, \citenamefont {Rubio}, \citenamefont {Shi}, \citenamefont
  {Tamii}, \citenamefont {Tashenov}, \citenamefont {Valiente-Dob\'on},
  \citenamefont {Walker}, \citenamefont {Woods},\ and\ \citenamefont
  {Xu}}]{alkhomashi2009}%
  \BibitemOpen
  \bibfield  {author} {\bibinfo {author} {\bibfnamefont {N.}~\bibnamefont
  {Alkhomashi}}, \bibinfo {author} {\bibfnamefont {P.~H.}\ \bibnamefont
  {Regan}}, \bibinfo {author} {\bibfnamefont {Z.}~\bibnamefont {Podoly\'ak}},
  \bibinfo {author} {\bibfnamefont {S.}~\bibnamefont {Pietri}}, \bibinfo
  {author} {\bibfnamefont {A.~B.}\ \bibnamefont {Garnsworthy}}, \bibinfo
  {author} {\bibfnamefont {S.~J.}\ \bibnamefont {Steer}}, \bibinfo {author}
  {\bibfnamefont {J.}~\bibnamefont {Benlliure}}, \bibinfo {author}
  {\bibfnamefont {E.}~\bibnamefont {Caserejos}}, \bibinfo {author}
  {\bibfnamefont {R.~F.}\ \bibnamefont {Casten}}, \bibinfo {author}
  {\bibfnamefont {J.}~\bibnamefont {Gerl}}, \bibinfo {author} {\bibfnamefont
  {H.~J.}\ \bibnamefont {Wollersheim}}, \bibinfo {author} {\bibfnamefont
  {J.}~\bibnamefont {Grebosz}}, \bibinfo {author} {\bibfnamefont
  {G.}~\bibnamefont {Farrelly}}, \bibinfo {author} {\bibfnamefont
  {M.}~\bibnamefont {G\'orska}}, \bibinfo {author} {\bibfnamefont
  {I.}~\bibnamefont {Kojouharov}}, \bibinfo {author} {\bibfnamefont
  {H.}~\bibnamefont {Schaffner}}, \bibinfo {author} {\bibfnamefont
  {A.}~\bibnamefont {Algora}}, \bibinfo {author} {\bibfnamefont
  {G.}~\bibnamefont {Benzoni}}, \bibinfo {author} {\bibfnamefont
  {A.}~\bibnamefont {Blazhev}}, \bibinfo {author} {\bibfnamefont
  {P.}~\bibnamefont {Boutachkov}}, \bibinfo {author} {\bibfnamefont {A.~M.}\
  \bibnamefont {Bruce}}, \bibinfo {author} {\bibfnamefont {A.~M.}\ \bibnamefont
  {Denis~Bacelar}}, \bibinfo {author} {\bibfnamefont {I.~J.}\ \bibnamefont
  {Cullen}}, \bibinfo {author} {\bibfnamefont {L.}~\bibnamefont {C\'aceres}},
  \bibinfo {author} {\bibfnamefont {P.}~\bibnamefont {Doornenbal}}, \bibinfo
  {author} {\bibfnamefont {M.~E.}\ \bibnamefont {Estevez}}, \bibinfo {author}
  {\bibfnamefont {Y.}~\bibnamefont {Fujita}}, \bibinfo {author} {\bibfnamefont
  {W.}~\bibnamefont {Gelletly}}, \bibinfo {author} {\bibfnamefont
  {R.}~\bibnamefont {Hoischen}}, \bibinfo {author} {\bibfnamefont
  {R.}~\bibnamefont {Kumar}}, \bibinfo {author} {\bibfnamefont
  {N.}~\bibnamefont {Kurz}}, \bibinfo {author} {\bibfnamefont {S.}~\bibnamefont
  {Lalkovski}}, \bibinfo {author} {\bibfnamefont {Z.}~\bibnamefont {Liu}},
  \bibinfo {author} {\bibfnamefont {C.}~\bibnamefont {Mihai}}, \bibinfo
  {author} {\bibfnamefont {F.}~\bibnamefont {Molina}}, \bibinfo {author}
  {\bibfnamefont {A.~I.}\ \bibnamefont {Morales}}, \bibinfo {author}
  {\bibfnamefont {D.}~\bibnamefont {M\"ucher}}, \bibinfo {author}
  {\bibfnamefont {W.}~\bibnamefont {Prokopowicz}}, \bibinfo {author}
  {\bibfnamefont {B.}~\bibnamefont {Rubio}}, \bibinfo {author} {\bibfnamefont
  {Y.}~\bibnamefont {Shi}}, \bibinfo {author} {\bibfnamefont {A.}~\bibnamefont
  {Tamii}}, \bibinfo {author} {\bibfnamefont {S.}~\bibnamefont {Tashenov}},
  \bibinfo {author} {\bibfnamefont {J.~J.}\ \bibnamefont {Valiente-Dob\'on}},
  \bibinfo {author} {\bibfnamefont {P.~M.}\ \bibnamefont {Walker}}, \bibinfo
  {author} {\bibfnamefont {P.~J.}\ \bibnamefont {Woods}}, \ and\ \bibinfo
  {author} {\bibfnamefont {F.~R.}\ \bibnamefont {Xu}},\ }\href {\doibase
  10.1103/PhysRevC.80.064308} {\bibfield  {journal} {\bibinfo  {journal} {Phys.
  Rev. C}\ }\textbf {\bibinfo {volume} {80}},\ \bibinfo {pages} {064308}
  (\bibinfo {year} {2009})}\BibitemShut {NoStop}%
\bibitem [{\citenamefont {Rudigier}\ \emph {et~al.}(2010)\citenamefont
  {Rudigier}, \citenamefont {R\'{e}gis}, \citenamefont {Jolie}, \citenamefont
  {Zell},\ and\ \citenamefont {Fransen}}]{rudigier2010}%
  \BibitemOpen
  \bibfield  {author} {\bibinfo {author} {\bibfnamefont {M.}~\bibnamefont
  {Rudigier}}, \bibinfo {author} {\bibfnamefont {J.-M.}\ \bibnamefont
  {R\'{e}gis}}, \bibinfo {author} {\bibfnamefont {J.}~\bibnamefont {Jolie}},
  \bibinfo {author} {\bibfnamefont {K.}~\bibnamefont {Zell}}, \ and\ \bibinfo
  {author} {\bibfnamefont {C.}~\bibnamefont {Fransen}},\ }\href {\doibase
  https://doi.org/10.1016/j.nuclphysa.2010.07.003} {\bibfield  {journal}
  {\bibinfo  {journal} {Nuclear Physics A}\ }\textbf {\bibinfo {volume}
  {847}},\ \bibinfo {pages} {89} (\bibinfo {year} {2010})}\BibitemShut
  {NoStop}%
\bibitem [{\citenamefont {Mason}\ \emph {et~al.}(2013)\citenamefont {Mason},
  \citenamefont {Podoly\'ak}, \citenamefont {M\ifmmode~\u{a}\else
  \u{a}\fi{}rginean}, \citenamefont {Regan}, \citenamefont {Stevenson},
  \citenamefont {Werner}, \citenamefont {Alexander}, \citenamefont {Algora},
  \citenamefont {Alharbi}, \citenamefont {Bowry}, \citenamefont {Britton},
  \citenamefont {Bruce}, \citenamefont {Bucurescu}, \citenamefont {Bunce},
  \citenamefont {C\u{a}ta-Danil}, \citenamefont {C\u{a}ta-Danil},
  \citenamefont {Cooper}, \citenamefont {Deleanu}, \citenamefont {Delion},
  \citenamefont {Filipescu}, \citenamefont {Gelletly}, \citenamefont
  {Ghi\ifmmode \mbox{\c{t}}\else \c{t}\fi{}\ifmmode~\u{a}\else \u{a}\fi{}},
  \citenamefont {Gheorghe}, \citenamefont {Glodariu}, \citenamefont {Ilie},
  \citenamefont {Ivanova}, \citenamefont {Kisyov}, \citenamefont {Lalkovski},
  \citenamefont {Lica}, \citenamefont {Liddick}, \citenamefont
  {M\ifmmode~\u{a}\else \u{a}\fi{}rginean}, \citenamefont {Mihai},
  \citenamefont {Mulholland}, \citenamefont {Nita}, \citenamefont {Negret},
  \citenamefont {Pascu}, \citenamefont {Rice}, \citenamefont {Roberts},
  \citenamefont {Sava}, \citenamefont {Smith}, \citenamefont {S\"oderstr\"om},
  \citenamefont {Stroe}, \citenamefont {Suliman}, \citenamefont {Suvaila},
  \citenamefont {Toma}, \citenamefont {Townsley}, \citenamefont {Wilson},
  \citenamefont {Wood}, \citenamefont {Zhekova},\ and\ \citenamefont
  {Zhou}}]{mason2013}%
  \BibitemOpen
  \bibfield  {author} {\bibinfo {author} {\bibfnamefont {P.~J.~R.}\
  \bibnamefont {Mason}}, \bibinfo {author} {\bibfnamefont {Z.}~\bibnamefont
  {Podoly\'ak}}, \bibinfo {author} {\bibfnamefont {N.}~\bibnamefont
  {M\ifmmode~\u{a}\else \u{a}\fi{}rginean}}, \bibinfo {author} {\bibfnamefont
  {P.~H.}\ \bibnamefont {Regan}}, \bibinfo {author} {\bibfnamefont {P.~D.}\
  \bibnamefont {Stevenson}}, \bibinfo {author} {\bibfnamefont {V.}~\bibnamefont
  {Werner}}, \bibinfo {author} {\bibfnamefont {T.}~\bibnamefont {Alexander}},
  \bibinfo {author} {\bibfnamefont {A.}~\bibnamefont {Algora}}, \bibinfo
  {author} {\bibfnamefont {T.}~\bibnamefont {Alharbi}}, \bibinfo {author}
  {\bibfnamefont {M.}~\bibnamefont {Bowry}}, \bibinfo {author} {\bibfnamefont
  {R.}~\bibnamefont {Britton}}, \bibinfo {author} {\bibfnamefont {A.~M.}\
  \bibnamefont {Bruce}}, \bibinfo {author} {\bibfnamefont {D.}~\bibnamefont
  {Bucurescu}}, \bibinfo {author} {\bibfnamefont {M.}~\bibnamefont {Bunce}},
  \bibinfo {author} {\bibfnamefont {G.}~\bibnamefont {C\u{a}ta-Danil}},
  \bibinfo {author} {\bibfnamefont {I.}~\bibnamefont {C\u{a}ta-Danil}},
  \bibinfo {author} {\bibfnamefont {N.}~\bibnamefont {Cooper}}, \bibinfo
  {author} {\bibfnamefont {D.}~\bibnamefont {Deleanu}}, \bibinfo {author}
  {\bibfnamefont {D.}~\bibnamefont {Delion}}, \bibinfo {author} {\bibfnamefont
  {D.}~\bibnamefont {Filipescu}}, \bibinfo {author} {\bibfnamefont
  {W.}~\bibnamefont {Gelletly}}, \bibinfo {author} {\bibfnamefont
  {D.}~\bibnamefont {Ghi\ifmmode \mbox{\c{t}}\else
  \c{t}\fi{}\ifmmode~\u{a}\else \u{a}\fi{}}}, \bibinfo {author} {\bibfnamefont
  {I.}~\bibnamefont {Gheorghe}}, \bibinfo {author} {\bibfnamefont
  {T.}~\bibnamefont {Glodariu}}, \bibinfo {author} {\bibfnamefont
  {G.}~\bibnamefont {Ilie}}, \bibinfo {author} {\bibfnamefont {D.}~\bibnamefont
  {Ivanova}}, \bibinfo {author} {\bibfnamefont {S.}~\bibnamefont {Kisyov}},
  \bibinfo {author} {\bibfnamefont {S.}~\bibnamefont {Lalkovski}}, \bibinfo
  {author} {\bibfnamefont {R.}~\bibnamefont {Lica}}, \bibinfo {author}
  {\bibfnamefont {S.~N.}\ \bibnamefont {Liddick}}, \bibinfo {author}
  {\bibfnamefont {R.}~\bibnamefont {M\ifmmode~\u{a}\else \u{a}\fi{}rginean}},
  \bibinfo {author} {\bibfnamefont {C.}~\bibnamefont {Mihai}}, \bibinfo
  {author} {\bibfnamefont {K.}~\bibnamefont {Mulholland}}, \bibinfo {author}
  {\bibfnamefont {C.~R.}\ \bibnamefont {Nita}}, \bibinfo {author}
  {\bibfnamefont {A.}~\bibnamefont {Negret}}, \bibinfo {author} {\bibfnamefont
  {S.}~\bibnamefont {Pascu}}, \bibinfo {author} {\bibfnamefont
  {S.}~\bibnamefont {Rice}}, \bibinfo {author} {\bibfnamefont {O.~J.}\
  \bibnamefont {Roberts}}, \bibinfo {author} {\bibfnamefont {T.}~\bibnamefont
  {Sava}}, \bibinfo {author} {\bibfnamefont {J.~F.}\ \bibnamefont {Smith}},
  \bibinfo {author} {\bibfnamefont {P.-A.}\ \bibnamefont {S\"oderstr\"om}},
  \bibinfo {author} {\bibfnamefont {L.}~\bibnamefont {Stroe}}, \bibinfo
  {author} {\bibfnamefont {G.}~\bibnamefont {Suliman}}, \bibinfo {author}
  {\bibfnamefont {R.}~\bibnamefont {Suvaila}}, \bibinfo {author} {\bibfnamefont
  {S.}~\bibnamefont {Toma}}, \bibinfo {author} {\bibfnamefont {C.}~\bibnamefont
  {Townsley}}, \bibinfo {author} {\bibfnamefont {E.}~\bibnamefont {Wilson}},
  \bibinfo {author} {\bibfnamefont {R.~T.}\ \bibnamefont {Wood}}, \bibinfo
  {author} {\bibfnamefont {M.}~\bibnamefont {Zhekova}}, \ and\ \bibinfo
  {author} {\bibfnamefont {C.}~\bibnamefont {Zhou}},\ }\href {\doibase
  10.1103/PhysRevC.88.044301} {\bibfield  {journal} {\bibinfo  {journal} {Phys.
  Rev. C}\ }\textbf {\bibinfo {volume} {88}},\ \bibinfo {pages} {044301}
  (\bibinfo {year} {2013})}\BibitemShut {NoStop}%
\bibitem [{\citenamefont {McGowan}\ \emph {et~al.}(1977)\citenamefont
  {McGowan}, \citenamefont {Milner}, \citenamefont {Sayer}, \citenamefont
  {Robinson},\ and\ \citenamefont {Stelson}}]{mcgowan1977}%
  \BibitemOpen
  \bibfield  {author} {\bibinfo {author} {\bibfnamefont {F.~K.}\ \bibnamefont
  {McGowan}}, \bibinfo {author} {\bibfnamefont {W.~T.}\ \bibnamefont {Milner}},
  \bibinfo {author} {\bibfnamefont {R.~O.}\ \bibnamefont {Sayer}}, \bibinfo
  {author} {\bibfnamefont {R.~H.}\ \bibnamefont {Robinson}}, \ and\ \bibinfo
  {author} {\bibfnamefont {P.~H.}\ \bibnamefont {Stelson}},\ }\href {\doibase
  https://doi.org/10.1016/0375-9474(77)90532-2} {\bibfield  {journal} {\bibinfo
   {journal} {Nucl. Phys. A}\ }\textbf {\bibinfo {volume} {289}},\ \bibinfo
  {pages} {253} (\bibinfo {year} {1977})}\BibitemShut {NoStop}%
\bibitem [{\citenamefont {Milner}\ \emph {et~al.}(1971)\citenamefont {Milner},
  \citenamefont {McGowan}, \citenamefont {Robinson}, \citenamefont {Stelson},\
  and\ \citenamefont {Sayer}}]{milner1971}%
  \BibitemOpen
  \bibfield  {author} {\bibinfo {author} {\bibfnamefont {W.~T.}\ \bibnamefont
  {Milner}}, \bibinfo {author} {\bibfnamefont {F.~K.}\ \bibnamefont {McGowan}},
  \bibinfo {author} {\bibfnamefont {R.~L.}\ \bibnamefont {Robinson}}, \bibinfo
  {author} {\bibfnamefont {P.~H.}\ \bibnamefont {Stelson}}, \ and\ \bibinfo
  {author} {\bibfnamefont {R.~O.}\ \bibnamefont {Sayer}},\ }\href {\doibase
  https://doi.org/10.1016/0375-9474(71)90161-8} {\bibfield  {journal} {\bibinfo
   {journal} {Nucl. Phys. A}\ }\textbf {\bibinfo {volume} {177}},\ \bibinfo
  {pages} {1} (\bibinfo {year} {1971})}\BibitemShut {NoStop}%
\bibitem [{\citenamefont {Kulessa}\ \emph {et~al.}(1989)\citenamefont
  {Kulessa}, \citenamefont {Bengtsson}, \citenamefont {Bohn}, \citenamefont
  {Emling}, \citenamefont {Faestermann}, \citenamefont {{Von Feilitsch}},
  \citenamefont {Grosse}, \citenamefont {Nazarewicz}, \citenamefont {Schwalm},\
  and\ \citenamefont {Wollersheim}}]{kulessa1989}%
  \BibitemOpen
  \bibfield  {author} {\bibinfo {author} {\bibfnamefont {R.}~\bibnamefont
  {Kulessa}}, \bibinfo {author} {\bibfnamefont {R.}~\bibnamefont {Bengtsson}},
  \bibinfo {author} {\bibfnamefont {H.}~\bibnamefont {Bohn}}, \bibinfo {author}
  {\bibfnamefont {H.}~\bibnamefont {Emling}}, \bibinfo {author} {\bibfnamefont
  {T.}~\bibnamefont {Faestermann}}, \bibinfo {author} {\bibfnamefont
  {F.}~\bibnamefont {{Von Feilitsch}}}, \bibinfo {author} {\bibfnamefont
  {E.}~\bibnamefont {Grosse}}, \bibinfo {author} {\bibfnamefont
  {W.}~\bibnamefont {Nazarewicz}}, \bibinfo {author} {\bibfnamefont
  {D.}~\bibnamefont {Schwalm}}, \ and\ \bibinfo {author} {\bibfnamefont
  {H.~J.}\ \bibnamefont {Wollersheim}},\ }\href {\doibase
  https://doi.org/10.1016/0370-2693(89)91439-1} {\bibfield  {journal} {\bibinfo
   {journal} {Phys. Lett. B}\ }\textbf {\bibinfo {volume} {218}},\ \bibinfo
  {pages} {421} (\bibinfo {year} {1989})}\BibitemShut {NoStop}%
\bibitem [{\citenamefont {Monnand}\ \emph {et~al.}(1969)\citenamefont
  {Monnand}, \citenamefont {Blachot},\ and\ \citenamefont
  {Moussa}}]{monnand1969}%
  \BibitemOpen
  \bibfield  {author} {\bibinfo {author} {\bibfnamefont {E.}~\bibnamefont
  {Monnand}}, \bibinfo {author} {\bibfnamefont {J.}~\bibnamefont {Blachot}}, \
  and\ \bibinfo {author} {\bibfnamefont {A.}~\bibnamefont {Moussa}},\ }\href
  {\doibase https://doi.org/10.1016/0375-9474(69)91055-0} {\bibfield  {journal}
  {\bibinfo  {journal} {Nucl. Phys. A}\ }\textbf {\bibinfo {volume} {134}},\
  \bibinfo {pages} {321} (\bibinfo {year} {1969})}\BibitemShut {NoStop}%
\bibitem [{\citenamefont {Ngijoi-Yogo}\ \emph {et~al.}(2007)\citenamefont
  {Ngijoi-Yogo}, \citenamefont {Tandel}, \citenamefont {Mukherjee},
  \citenamefont {Shestakova}, \citenamefont {Chowdhury}, \citenamefont {Wu},
  \citenamefont {Cline}, \citenamefont {Hayes}, \citenamefont {Teng},
  \citenamefont {Clark}, \citenamefont {Fallon}, \citenamefont {Macchiavelli},
  \citenamefont {Vetter}, \citenamefont {Kondev}, \citenamefont {Langdown},
  \citenamefont {Walker}, \citenamefont {Wheldon},\ and\ \citenamefont
  {Cullen}}]{ngijoi-yogo2007}%
  \BibitemOpen
  \bibfield  {author} {\bibinfo {author} {\bibfnamefont {E.}~\bibnamefont
  {Ngijoi-Yogo}}, \bibinfo {author} {\bibfnamefont {S.~K.}\ \bibnamefont
  {Tandel}}, \bibinfo {author} {\bibfnamefont {G.}~\bibnamefont {Mukherjee}},
  \bibinfo {author} {\bibfnamefont {I.}~\bibnamefont {Shestakova}}, \bibinfo
  {author} {\bibfnamefont {P.}~\bibnamefont {Chowdhury}}, \bibinfo {author}
  {\bibfnamefont {C.~Y.}\ \bibnamefont {Wu}}, \bibinfo {author} {\bibfnamefont
  {D.}~\bibnamefont {Cline}}, \bibinfo {author} {\bibfnamefont {A.~B.}\
  \bibnamefont {Hayes}}, \bibinfo {author} {\bibfnamefont {R.}~\bibnamefont
  {Teng}}, \bibinfo {author} {\bibfnamefont {R.~M.}\ \bibnamefont {Clark}},
  \bibinfo {author} {\bibfnamefont {P.}~\bibnamefont {Fallon}}, \bibinfo
  {author} {\bibfnamefont {A.~O.}\ \bibnamefont {Macchiavelli}}, \bibinfo
  {author} {\bibfnamefont {K.}~\bibnamefont {Vetter}}, \bibinfo {author}
  {\bibfnamefont {F.~G.}\ \bibnamefont {Kondev}}, \bibinfo {author}
  {\bibfnamefont {S.}~\bibnamefont {Langdown}}, \bibinfo {author}
  {\bibfnamefont {P.~M.}\ \bibnamefont {Walker}}, \bibinfo {author}
  {\bibfnamefont {C.}~\bibnamefont {Wheldon}}, \ and\ \bibinfo {author}
  {\bibfnamefont {D.~M.}\ \bibnamefont {Cullen}},\ }\href {\doibase
  10.1103/PhysRevC.75.034305} {\bibfield  {journal} {\bibinfo  {journal} {Phys.
  Rev. C}\ }\textbf {\bibinfo {volume} {75}},\ \bibinfo {pages} {034305}
  (\bibinfo {year} {2007})}\BibitemShut {NoStop}%
\bibitem [{\citenamefont {Tandel}\ \emph
  {et~al.}(2008{\natexlab{a}})\citenamefont {Tandel}, \citenamefont {Tandel},
  \citenamefont {Chowdhury}, \citenamefont {Cline}, \citenamefont {Wu},
  \citenamefont {Carpenter}, \citenamefont {Janssens}, \citenamefont {Khoo},
  \citenamefont {Lauritsen}, \citenamefont {Lister}, \citenamefont
  {Seweryniak},\ and\ \citenamefont {Zhu}}]{tandel2008-1}%
  \BibitemOpen
  \bibfield  {author} {\bibinfo {author} {\bibfnamefont {U.~S.}\ \bibnamefont
  {Tandel}}, \bibinfo {author} {\bibfnamefont {S.~K.}\ \bibnamefont {Tandel}},
  \bibinfo {author} {\bibfnamefont {P.}~\bibnamefont {Chowdhury}}, \bibinfo
  {author} {\bibfnamefont {D.}~\bibnamefont {Cline}}, \bibinfo {author}
  {\bibfnamefont {C.~Y.}\ \bibnamefont {Wu}}, \bibinfo {author} {\bibfnamefont
  {M.~P.}\ \bibnamefont {Carpenter}}, \bibinfo {author} {\bibfnamefont
  {R.~V.~F.}\ \bibnamefont {Janssens}}, \bibinfo {author} {\bibfnamefont
  {T.~L.}\ \bibnamefont {Khoo}}, \bibinfo {author} {\bibfnamefont
  {T.}~\bibnamefont {Lauritsen}}, \bibinfo {author} {\bibfnamefont {C.~J.}\
  \bibnamefont {Lister}}, \bibinfo {author} {\bibfnamefont {D.}~\bibnamefont
  {Seweryniak}}, \ and\ \bibinfo {author} {\bibfnamefont {S.}~\bibnamefont
  {Zhu}},\ }\href {\doibase 10.1103/PhysRevLett.101.182503} {\bibfield
  {journal} {\bibinfo  {journal} {Phys. Rev. Lett.}\ }\textbf {\bibinfo
  {volume} {101}},\ \bibinfo {pages} {182503} (\bibinfo {year}
  {2008}{\natexlab{a}})}\BibitemShut {NoStop}%
\bibitem [{\citenamefont {Broda}(2006)}]{broda2006}%
  \BibitemOpen
  \bibfield  {author} {\bibinfo {author} {\bibfnamefont {R.}~\bibnamefont
  {Broda}},\ }\href {\doibase 10.1088/0954-3899/32/6/r01} {\bibfield  {journal}
  {\bibinfo  {journal} {J. Phys. G}\ }\textbf {\bibinfo {volume} {32}},\
  \bibinfo {pages} {R151} (\bibinfo {year} {2006})}\BibitemShut {NoStop}%
\bibitem [{\citenamefont {Broda}\ \emph {et~al.}(2012)\citenamefont {Broda},
  \citenamefont {Paw\l{}at}, \citenamefont {Kr\'olas}, \citenamefont
  {Janssens}, \citenamefont {Zhu}, \citenamefont {Walters}, \citenamefont
  {Fornal}, \citenamefont {Chiara}, \citenamefont {Carpenter}, \citenamefont
  {Hoteling}, \citenamefont {Iskra}, \citenamefont {Kondev}, \citenamefont
  {Lauritsen}, \citenamefont {Seweryniak}, \citenamefont {Stefanescu},
  \citenamefont {Wang},\ and\ \citenamefont {Wrzesi\ifmmode~\acute{n}\else
  \'{n}\fi{}ski}}]{broda2012}%
  \BibitemOpen
  \bibfield  {author} {\bibinfo {author} {\bibfnamefont {R.}~\bibnamefont
  {Broda}}, \bibinfo {author} {\bibfnamefont {T.}~\bibnamefont {Paw\l{}at}},
  \bibinfo {author} {\bibfnamefont {W.}~\bibnamefont {Kr\'olas}}, \bibinfo
  {author} {\bibfnamefont {R.~V.~F.}\ \bibnamefont {Janssens}}, \bibinfo
  {author} {\bibfnamefont {S.}~\bibnamefont {Zhu}}, \bibinfo {author}
  {\bibfnamefont {W.~B.}\ \bibnamefont {Walters}}, \bibinfo {author}
  {\bibfnamefont {B.}~\bibnamefont {Fornal}}, \bibinfo {author} {\bibfnamefont
  {C.~J.}\ \bibnamefont {Chiara}}, \bibinfo {author} {\bibfnamefont {M.~P.}\
  \bibnamefont {Carpenter}}, \bibinfo {author} {\bibfnamefont {N.}~\bibnamefont
  {Hoteling}}, \bibinfo {author} {\bibfnamefont {L.~W.}\ \bibnamefont {Iskra}},
  \bibinfo {author} {\bibfnamefont {F.~G.}\ \bibnamefont {Kondev}}, \bibinfo
  {author} {\bibfnamefont {T.}~\bibnamefont {Lauritsen}}, \bibinfo {author}
  {\bibfnamefont {D.}~\bibnamefont {Seweryniak}}, \bibinfo {author}
  {\bibfnamefont {I.}~\bibnamefont {Stefanescu}}, \bibinfo {author}
  {\bibfnamefont {X.}~\bibnamefont {Wang}}, \ and\ \bibinfo {author}
  {\bibfnamefont {J.}~\bibnamefont {Wrzesi\ifmmode~\acute{n}\else
  \'{n}\fi{}ski}},\ }\href {\doibase 10.1103/PhysRevC.86.064312} {\bibfield
  {journal} {\bibinfo  {journal} {Phys. Rev. C}\ }\textbf {\bibinfo {volume}
  {86}},\ \bibinfo {pages} {064312} (\bibinfo {year} {2012})}\BibitemShut
  {NoStop}%
\bibitem [{\citenamefont {Prasher}(2015)}]{prasher2015}%
  \BibitemOpen
  \bibfield  {author} {\bibinfo {author} {\bibfnamefont {V.~S.}\ \bibnamefont
  {Prasher}},\ }\href@noop {} {Ph.D. thesis},\ \bibinfo  {school} {University
  of Massachusetts Lowell} (\bibinfo {year} {2015})\BibitemShut {NoStop}%
\bibitem [{\citenamefont {Wu}\ \emph {et~al.}(2016)\citenamefont {Wu},
  \citenamefont {Cline}, \citenamefont {Hayes}, \citenamefont {Flight},
  \citenamefont {Melchionna}, \citenamefont {Zhou}, \citenamefont {Lee},
  \citenamefont {Swan}, \citenamefont {Fox},\ and\ \citenamefont
  {Anderson}}]{wu2016}%
  \BibitemOpen
  \bibfield  {author} {\bibinfo {author} {\bibfnamefont {C.~Y.}\ \bibnamefont
  {Wu}}, \bibinfo {author} {\bibfnamefont {D.}~\bibnamefont {Cline}}, \bibinfo
  {author} {\bibfnamefont {A.}~\bibnamefont {Hayes}}, \bibinfo {author}
  {\bibfnamefont {R.~S.}\ \bibnamefont {Flight}}, \bibinfo {author}
  {\bibfnamefont {A.~M.}\ \bibnamefont {Melchionna}}, \bibinfo {author}
  {\bibfnamefont {C.}~\bibnamefont {Zhou}}, \bibinfo {author} {\bibfnamefont
  {I.~Y.}\ \bibnamefont {Lee}}, \bibinfo {author} {\bibfnamefont
  {D.}~\bibnamefont {Swan}}, \bibinfo {author} {\bibfnamefont {R.}~\bibnamefont
  {Fox}}, \ and\ \bibinfo {author} {\bibfnamefont {J.~T.}\ \bibnamefont
  {Anderson}},\ }\href {\doibase https://doi.org/10.1016/j.nima.2016.01.034}
  {\bibfield  {journal} {\bibinfo  {journal} {Nucl. Instrum. Meth. A}\ }\textbf
  {\bibinfo {volume} {814}},\ \bibinfo {pages} {6} (\bibinfo {year}
  {2016})}\BibitemShut {NoStop}%
\bibitem [{dgs()}]{dgssort}%
  \BibitemOpen
  \href@noop {} {}\bibinfo {howpublished}
  {\url{https://www.phy.anl.gov/gammasphere/doc/DGSSort/}}\BibitemShut
  {NoStop}%
\bibitem [{\citenamefont {Brun}\ and\ \citenamefont
  {Rademakers}(1997)}]{brun1997}%
  \BibitemOpen
  \bibfield  {author} {\bibinfo {author} {\bibfnamefont {R.}~\bibnamefont
  {Brun}}\ and\ \bibinfo {author} {\bibfnamefont {F.}~\bibnamefont
  {Rademakers}},\ }\href {\doibase
  https://doi.org/10.1016/S0168-9002(97)00048-X} {\bibfield  {journal}
  {\bibinfo  {journal} {Nucl. Instrum. Meth. A}\ }\textbf {\bibinfo {volume}
  {389}},\ \bibinfo {pages} {81} (\bibinfo {year} {1997})},\ \bibinfo {note}
  {new Computing Techniques in Physics Research V}\BibitemShut {NoStop}%
\bibitem [{\citenamefont {Radford}(1995)}]{radware}%
  \BibitemOpen
  \bibfield  {author} {\bibinfo {author} {\bibfnamefont {D.~C.}\ \bibnamefont
  {Radford}},\ }\href {\doibase https://doi.org/10.1016/0168-9002(95)00183-2}
  {\bibfield  {journal} {\bibinfo  {journal} {Nucl. Instrum. Meth. A}\ }\textbf
  {\bibinfo {volume} {361}},\ \bibinfo {pages} {297} (\bibinfo {year}
  {1995})}\BibitemShut {NoStop}%
\bibitem [{\citenamefont {Sonzogni}(2002)}]{A=136}%
  \BibitemOpen
  \bibfield  {author} {\bibinfo {author} {\bibfnamefont {A.~A.}\ \bibnamefont
  {Sonzogni}},\ }\href {\doibase https://doi.org/10.1006/ndsh.2002.0008}
  {\bibfield  {journal} {\bibinfo  {journal} {Nucl. Data Sheets}\ }\textbf
  {\bibinfo {volume} {95}},\ \bibinfo {pages} {837} (\bibinfo {year}
  {2002})}\BibitemShut {NoStop}%
\bibitem [{\citenamefont {Krämer-Flecken}\ \emph {et~al.}(1989)\citenamefont
  {Krämer-Flecken}, \citenamefont {Morek}, \citenamefont {Lieder},
  \citenamefont {Gast}, \citenamefont {Hebbinghaus}, \citenamefont {Jäger},\
  and\ \citenamefont {Urban}}]{kramerflecken1989}%
  \BibitemOpen
  \bibfield  {author} {\bibinfo {author} {\bibfnamefont {A.}~\bibnamefont
  {Kr\"{a}mer-Flecken}}, \bibinfo {author} {\bibfnamefont {T.}~\bibnamefont
  {Morek}}, \bibinfo {author} {\bibfnamefont {R.~M.}\ \bibnamefont {Lieder}},
  \bibinfo {author} {\bibfnamefont {W.}~\bibnamefont {Gast}}, \bibinfo {author}
  {\bibfnamefont {G.}~\bibnamefont {Hebbinghaus}}, \bibinfo {author}
  {\bibfnamefont {H.~M.}\ \bibnamefont {J\"{a}ger}}, \ and\ \bibinfo {author}
  {\bibfnamefont {W.}~\bibnamefont {Urban}},\ }\href {\doibase
  https://doi.org/10.1016/0168-9002(89)90706-7} {\bibfield  {journal} {\bibinfo
   {journal} {Nucl. Instrum. Meth. A}\ }\textbf {\bibinfo {volume} {275}},\
  \bibinfo {pages} {333} (\bibinfo {year} {1989})}\BibitemShut {NoStop}%
\bibitem [{\citenamefont {Hilton}\ and\ \citenamefont
  {Mang}(1979)}]{hilton1979}%
  \BibitemOpen
  \bibfield  {author} {\bibinfo {author} {\bibfnamefont {R.~R.}\ \bibnamefont
  {Hilton}}\ and\ \bibinfo {author} {\bibfnamefont {H.~J.}\ \bibnamefont
  {Mang}},\ }\href {\doibase 10.1103/PhysRevLett.43.1979} {\bibfield  {journal}
  {\bibinfo  {journal} {Phys. Rev. Lett.}\ }\textbf {\bibinfo {volume} {43}},\
  \bibinfo {pages} {1979} (\bibinfo {year} {1979})}\BibitemShut {NoStop}%
\bibitem [{\citenamefont {Davydov}\ and\ \citenamefont
  {Filippov}(1958)}]{davydov1958}%
  \BibitemOpen
  \bibfield  {author} {\bibinfo {author} {\bibfnamefont {A.}~\bibnamefont
  {Davydov}}\ and\ \bibinfo {author} {\bibfnamefont {G.}~\bibnamefont
  {Filippov}},\ }\href {\doibase https://doi.org/10.1016/0029-5582(58)90153-6}
  {\bibfield  {journal} {\bibinfo  {journal} {Nucl. Phys.}\ }\textbf {\bibinfo
  {volume} {8}},\ \bibinfo {pages} {237} (\bibinfo {year} {1958})}\BibitemShut
  {NoStop}%
\bibitem [{\citenamefont {Wu}\ and\ \citenamefont {Cline}(1996)}]{wu1996}%
  \BibitemOpen
  \bibfield  {author} {\bibinfo {author} {\bibfnamefont {C.~Y.}\ \bibnamefont
  {Wu}}\ and\ \bibinfo {author} {\bibfnamefont {D.}~\bibnamefont {Cline}},\
  }\href {\doibase 10.1103/PhysRevC.54.2356} {\bibfield  {journal} {\bibinfo
  {journal} {Phys. Rev. C}\ }\textbf {\bibinfo {volume} {54}},\ \bibinfo
  {pages} {2356} (\bibinfo {year} {1996})}\BibitemShut {NoStop}%
\bibitem [{\citenamefont {McCutchan}\ \emph {et~al.}(2007)\citenamefont
  {McCutchan}, \citenamefont {Bonatsos}, \citenamefont {Zamfir},\ and\
  \citenamefont {Casten}}]{mccutchan2007}%
  \BibitemOpen
  \bibfield  {author} {\bibinfo {author} {\bibfnamefont {E.~A.}\ \bibnamefont
  {McCutchan}}, \bibinfo {author} {\bibfnamefont {D.}~\bibnamefont {Bonatsos}},
  \bibinfo {author} {\bibfnamefont {N.~V.}\ \bibnamefont {Zamfir}}, \ and\
  \bibinfo {author} {\bibfnamefont {R.~F.}\ \bibnamefont {Casten}},\ }\href
  {\doibase 10.1103/PhysRevC.76.024306} {\bibfield  {journal} {\bibinfo
  {journal} {Phys. Rev. C}\ }\textbf {\bibinfo {volume} {76}},\ \bibinfo
  {pages} {024306} (\bibinfo {year} {2007})}\BibitemShut {NoStop}%
\bibitem [{\citenamefont {Zamfir}\ and\ \citenamefont
  {Casten}(1991)}]{zamfir1991}%
  \BibitemOpen
  \bibfield  {author} {\bibinfo {author} {\bibfnamefont {N.}~\bibnamefont
  {Zamfir}}\ and\ \bibinfo {author} {\bibfnamefont {R.}~\bibnamefont
  {Casten}},\ }\href {\doibase https://doi.org/10.1016/0370-2693(91)91610-8}
  {\bibfield  {journal} {\bibinfo  {journal} {Phys. Lett. B}\ }\textbf
  {\bibinfo {volume} {260}},\ \bibinfo {pages} {265} (\bibinfo {year}
  {1991})}\BibitemShut {NoStop}%
\bibitem [{\citenamefont {Moore}\ and\ \citenamefont
  {White}(1960)}]{moore1960}%
  \BibitemOpen
  \bibfield  {author} {\bibinfo {author} {\bibfnamefont {R.~B.}\ \bibnamefont
  {Moore}}\ and\ \bibinfo {author} {\bibfnamefont {W.}~\bibnamefont {White}},\
  }\href {\doibase 10.1139/p60-123} {\bibfield  {journal} {\bibinfo  {journal}
  {Canadian Journal of Physics}\ }\textbf {\bibinfo {volume} {38}},\ \bibinfo
  {pages} {1149} (\bibinfo {year} {1960})}\BibitemShut {NoStop}%
\bibitem [{\citenamefont {Wilets}\ and\ \citenamefont
  {Jean}(1956)}]{wilets1956}%
  \BibitemOpen
  \bibfield  {author} {\bibinfo {author} {\bibfnamefont {L.}~\bibnamefont
  {Wilets}}\ and\ \bibinfo {author} {\bibfnamefont {M.}~\bibnamefont {Jean}},\
  }\href {\doibase 10.1103/PhysRev.102.788} {\bibfield  {journal} {\bibinfo
  {journal} {Phys. Rev.}\ }\textbf {\bibinfo {volume} {102}},\ \bibinfo {pages}
  {788} (\bibinfo {year} {1956})}\BibitemShut {NoStop}%
\bibitem [{\citenamefont {Hayes}\ \emph {et~al.}(2002)\citenamefont {Hayes},
  \citenamefont {Cline}, \citenamefont {Wu}, \citenamefont {Simon},
  \citenamefont {Teng}, \citenamefont {Gerl}, \citenamefont {Schlegel},
  \citenamefont {Wollersheim}, \citenamefont {Macchiavelli}, \citenamefont
  {Vetter}, \citenamefont {Napiorkowski},\ and\ \citenamefont
  {Srebrny}}]{hayes2002}%
  \BibitemOpen
  \bibfield  {author} {\bibinfo {author} {\bibfnamefont {A.~B.}\ \bibnamefont
  {Hayes}}, \bibinfo {author} {\bibfnamefont {D.}~\bibnamefont {Cline}},
  \bibinfo {author} {\bibfnamefont {C.~Y.}\ \bibnamefont {Wu}}, \bibinfo
  {author} {\bibfnamefont {M.~W.}\ \bibnamefont {Simon}}, \bibinfo {author}
  {\bibfnamefont {R.}~\bibnamefont {Teng}}, \bibinfo {author} {\bibfnamefont
  {J.}~\bibnamefont {Gerl}}, \bibinfo {author} {\bibfnamefont {C.}~\bibnamefont
  {Schlegel}}, \bibinfo {author} {\bibfnamefont {H.~J.}\ \bibnamefont
  {Wollersheim}}, \bibinfo {author} {\bibfnamefont {A.~O.}\ \bibnamefont
  {Macchiavelli}}, \bibinfo {author} {\bibfnamefont {K.}~\bibnamefont
  {Vetter}}, \bibinfo {author} {\bibfnamefont {P.}~\bibnamefont
  {Napiorkowski}}, \ and\ \bibinfo {author} {\bibfnamefont {J.}~\bibnamefont
  {Srebrny}},\ }\href {\doibase 10.1103/PhysRevLett.89.242501} {\bibfield
  {journal} {\bibinfo  {journal} {Phys. Rev. Lett.}\ }\textbf {\bibinfo
  {volume} {89}},\ \bibinfo {pages} {242501} (\bibinfo {year}
  {2002})}\BibitemShut {NoStop}%
\bibitem [{\citenamefont {Tandel}\ \emph
  {et~al.}(2008{\natexlab{b}})\citenamefont {Tandel}, \citenamefont {Knox},
  \citenamefont {Parnell-Lampen}, \citenamefont {Tandel}, \citenamefont
  {Chowdhury}, \citenamefont {Carpenter}, \citenamefont {Janssens},
  \citenamefont {Khoo}, \citenamefont {Lauritsen}, \citenamefont {Lister},
  \citenamefont {Seweryniak}, \citenamefont {Wang}, \citenamefont {Zhu},
  \citenamefont {Hartley},\ and\ \citenamefont {Zhang}}]{tandel2008-2}%
  \BibitemOpen
  \bibfield  {author} {\bibinfo {author} {\bibfnamefont {S.~K.}\ \bibnamefont
  {Tandel}}, \bibinfo {author} {\bibfnamefont {A.~J.}\ \bibnamefont {Knox}},
  \bibinfo {author} {\bibfnamefont {C.}~\bibnamefont {Parnell-Lampen}},
  \bibinfo {author} {\bibfnamefont {U.~S.}\ \bibnamefont {Tandel}}, \bibinfo
  {author} {\bibfnamefont {P.}~\bibnamefont {Chowdhury}}, \bibinfo {author}
  {\bibfnamefont {M.~P.}\ \bibnamefont {Carpenter}}, \bibinfo {author}
  {\bibfnamefont {R.~V.~F.}\ \bibnamefont {Janssens}}, \bibinfo {author}
  {\bibfnamefont {T.~L.}\ \bibnamefont {Khoo}}, \bibinfo {author}
  {\bibfnamefont {T.}~\bibnamefont {Lauritsen}}, \bibinfo {author}
  {\bibfnamefont {C.~J.}\ \bibnamefont {Lister}}, \bibinfo {author}
  {\bibfnamefont {D.}~\bibnamefont {Seweryniak}}, \bibinfo {author}
  {\bibfnamefont {X.}~\bibnamefont {Wang}}, \bibinfo {author} {\bibfnamefont
  {S.}~\bibnamefont {Zhu}}, \bibinfo {author} {\bibfnamefont {D.~J.}\
  \bibnamefont {Hartley}}, \ and\ \bibinfo {author} {\bibfnamefont {J.-y.}\
  \bibnamefont {Zhang}},\ }\href {\doibase 10.1103/PhysRevC.77.024313}
  {\bibfield  {journal} {\bibinfo  {journal} {Phys. Rev. C}\ }\textbf {\bibinfo
  {volume} {77}},\ \bibinfo {pages} {024313} (\bibinfo {year}
  {2008}{\natexlab{b}})}\BibitemShut {NoStop}%
\bibitem [{\citenamefont {Wheldon}\ \emph {et~al.}(1999)\citenamefont
  {Wheldon}, \citenamefont {Walker}, \citenamefont {Regan}, \citenamefont
  {Saitoh}, \citenamefont {Hashimoto}, \citenamefont {Sletten},\ and\
  \citenamefont {Xu}}]{wheldon1999}%
  \BibitemOpen
  \bibfield  {author} {\bibinfo {author} {\bibfnamefont {C.}~\bibnamefont
  {Wheldon}}, \bibinfo {author} {\bibfnamefont {P.}~\bibnamefont {Walker}},
  \bibinfo {author} {\bibfnamefont {P.}~\bibnamefont {Regan}}, \bibinfo
  {author} {\bibfnamefont {T.}~\bibnamefont {Saitoh}}, \bibinfo {author}
  {\bibfnamefont {N.}~\bibnamefont {Hashimoto}}, \bibinfo {author}
  {\bibfnamefont {G.}~\bibnamefont {Sletten}}, \ and\ \bibinfo {author}
  {\bibfnamefont {F.}~\bibnamefont {Xu}},\ }\href {\doibase
  https://doi.org/10.1016/S0375-9474(99)00160-8} {\bibfield  {journal}
  {\bibinfo  {journal} {Nucl. Phys. A}\ }\textbf {\bibinfo {volume} {652}},\
  \bibinfo {pages} {103} (\bibinfo {year} {1999})}\BibitemShut {NoStop}%
\bibitem [{\citenamefont {Modamio}\ \emph {et~al.}(2009)\citenamefont
  {Modamio}, \citenamefont {Jungclaus}, \citenamefont {Podolyak}, \citenamefont
  {Shi}, \citenamefont {Xu}, \citenamefont {Algora}, \citenamefont {Bazzacco},
  \citenamefont {Escrig}, \citenamefont {Fraile}, \citenamefont {Lenzi},
  \citenamefont {Marginean}, \citenamefont {Martinez}, \citenamefont {Napoli},
  \citenamefont {Schwengner},\ and\ \citenamefont {Ur}}]{modamio2009}%
  \BibitemOpen
  \bibfield  {author} {\bibinfo {author} {\bibfnamefont {V.}~\bibnamefont
  {Modamio}}, \bibinfo {author} {\bibfnamefont {A.}~\bibnamefont {Jungclaus}},
  \bibinfo {author} {\bibfnamefont {Z.}~\bibnamefont {Podolyak}}, \bibinfo
  {author} {\bibfnamefont {Y.}~\bibnamefont {Shi}}, \bibinfo {author}
  {\bibfnamefont {F.~R.}\ \bibnamefont {Xu}}, \bibinfo {author} {\bibfnamefont
  {A.}~\bibnamefont {Algora}}, \bibinfo {author} {\bibfnamefont
  {D.}~\bibnamefont {Bazzacco}}, \bibinfo {author} {\bibfnamefont
  {D.}~\bibnamefont {Escrig}}, \bibinfo {author} {\bibfnamefont {L.~M.}\
  \bibnamefont {Fraile}}, \bibinfo {author} {\bibfnamefont {S.}~\bibnamefont
  {Lenzi}}, \bibinfo {author} {\bibfnamefont {N.}~\bibnamefont {Marginean}},
  \bibinfo {author} {\bibfnamefont {T.}~\bibnamefont {Martinez}}, \bibinfo
  {author} {\bibfnamefont {D.~R.}\ \bibnamefont {Napoli}}, \bibinfo {author}
  {\bibfnamefont {R.}~\bibnamefont {Schwengner}}, \ and\ \bibinfo {author}
  {\bibfnamefont {C.~A.}\ \bibnamefont {Ur}},\ }\href {\doibase
  10.1103/PhysRevC.79.024310} {\bibfield  {journal} {\bibinfo  {journal} {Phys.
  Rev. C}\ }\textbf {\bibinfo {volume} {79}},\ \bibinfo {pages} {024310}
  (\bibinfo {year} {2009})}\BibitemShut {NoStop}%
\bibitem [{\citenamefont {Heyde}\ and\ \citenamefont {Wood}(2011)}]{heyde2011}%
  \BibitemOpen
  \bibfield  {author} {\bibinfo {author} {\bibfnamefont {K.}~\bibnamefont
  {Heyde}}\ and\ \bibinfo {author} {\bibfnamefont {J.~L.}\ \bibnamefont
  {Wood}},\ }\href {\doibase 10.1103/RevModPhys.83.1467} {\bibfield  {journal}
  {\bibinfo  {journal} {Rev. Mod. Phys.}\ }\textbf {\bibinfo {volume} {83}},\
  \bibinfo {pages} {1467} (\bibinfo {year} {2011})}\BibitemShut {NoStop}%
\bibitem [{\citenamefont {rd}\ and\ \citenamefont
  {Vogel}(1969)}]{neergard1969}%
  \BibitemOpen
  \bibfield  {author} {\bibinfo {author} {\bibfnamefont {K.~N.}\ \bibnamefont
  {Neergård}}\ and\ \bibinfo {author} {\bibfnamefont {P.}~\bibnamefont {Vogel}},\
  }\href {\doibase https://doi.org/10.1016/0370-2693(69)90400-6} {\bibfield
  {journal} {\bibinfo  {journal} {Phys. Lett. B}\ }\textbf {\bibinfo {volume}
  {30}},\ \bibinfo {pages} {75} (\bibinfo {year} {1969})}\BibitemShut {NoStop}%
\bibitem [{\citenamefont {rd}\ and\ \citenamefont
  {Vogel}(1970)}]{neergard1970}%
  \BibitemOpen
  \bibfield  {author} {\bibinfo {author} {\bibfnamefont {K.~N.}\ \bibnamefont
  {Neergård}}\ and\ \bibinfo {author} {\bibfnamefont {P.}~\bibnamefont {Vogel}},\
  }\href {\doibase https://doi.org/10.1016/0375-9474(70)90388-X} {\bibfield
  {journal} {\bibinfo  {journal} {Nucl. Phys. A}\ }\textbf {\bibinfo {volume}
  {149}},\ \bibinfo {pages} {217} (\bibinfo {year} {1970})}\BibitemShut
  {NoStop}%
\bibitem [{\citenamefont {Vogel}(1976)}]{vogel1976}%
  \BibitemOpen
  \bibfield  {author} {\bibinfo {author} {\bibfnamefont {P.}~\bibnamefont
  {Vogel}},\ }\href {\doibase https://doi.org/10.1016/0370-2693(76)90699-7}
  {\bibfield  {journal} {\bibinfo  {journal} {Phys. Lett. B}\ }\textbf
  {\bibinfo {volume} {60}},\ \bibinfo {pages} {431} (\bibinfo {year}
  {1976})}\BibitemShut {NoStop}%
\bibitem [{\citenamefont {Toki}\ \emph {et~al.}(1977)\citenamefont {Toki},
  \citenamefont {Neergård}, \citenamefont {Vogel},\ and\ \citenamefont
  {Faessler}}]{toki1977}%
  \BibitemOpen
  \bibfield  {author} {\bibinfo {author} {\bibfnamefont {H.}~\bibnamefont
  {Toki}}, \bibinfo {author} {\bibfnamefont {K.}~\bibnamefont {Neerg\aa rd}},
  \bibinfo {author} {\bibfnamefont {P.}~\bibnamefont {Vogel}}, \ and\ \bibinfo
  {author} {\bibfnamefont {A.}~\bibnamefont {Faessler}},\ }\href {\doibase
  https://doi.org/10.1016/0375-9474(77)90417-1} {\bibfield  {journal} {\bibinfo
   {journal} {Nucl. Phys. A}\ }\textbf {\bibinfo {volume} {279}},\ \bibinfo
  {pages} {1} (\bibinfo {year} {1977})}\BibitemShut {NoStop}%
\bibitem [{\citenamefont {Cottle}\ and\ \citenamefont
  {Zamfir}(1996)}]{cottle1996}%
  \BibitemOpen
  \bibfield  {author} {\bibinfo {author} {\bibfnamefont {P.~D.}\ \bibnamefont
  {Cottle}}\ and\ \bibinfo {author} {\bibfnamefont {N.~V.}\ \bibnamefont
  {Zamfir}},\ }\href {\doibase 10.1103/PhysRevC.54.176} {\bibfield  {journal}
  {\bibinfo  {journal} {Phys. Rev. C}\ }\textbf {\bibinfo {volume} {54}},\
  \bibinfo {pages} {176} (\bibinfo {year} {1996})}\BibitemShut {NoStop}%
\bibitem [{\citenamefont {Alaga}\ \emph {et~al.}(1955)\citenamefont {Alaga},
  \citenamefont {Alder}, \citenamefont {Bohr},\ and\ \citenamefont
  {Mottelson}}]{alaga1955}%
  \BibitemOpen
  \bibfield  {author} {\bibinfo {author} {\bibfnamefont {G.}~\bibnamefont
  {Alaga}}, \bibinfo {author} {\bibfnamefont {K.}~\bibnamefont {Alder}},
  \bibinfo {author} {\bibfnamefont {A.}~\bibnamefont {Bohr}}, \ and\ \bibinfo
  {author} {\bibfnamefont {B.~R.}\ \bibnamefont {Mottelson}},\ }\href
  {http://gymarkiv.sdu.dk/MFM/kdvs/mfm\%2020-29/mfm-29-9.pdf} {\bibfield
  {journal} {\bibinfo  {journal} {Dan. Mat. Fys. Medd.}\ }\textbf {\bibinfo
  {volume} {29}},\ \bibinfo {pages} {1} (\bibinfo {year} {1955})}\BibitemShut
  {NoStop}%
\bibitem [{\citenamefont {Khoo}\ \emph {et~al.}(1973)\citenamefont {Khoo},
  \citenamefont {Waddington}, \citenamefont {Preibisz},\ and\ \citenamefont
  {Johns}}]{khoo1976}%
  \BibitemOpen
  \bibfield  {author} {\bibinfo {author} {\bibfnamefont {T.}~\bibnamefont
  {Khoo}}, \bibinfo {author} {\bibfnamefont {J.}~\bibnamefont {Waddington}},
  \bibinfo {author} {\bibfnamefont {Z.}~\bibnamefont {Preibisz}}, \ and\
  \bibinfo {author} {\bibfnamefont {M.}~\bibnamefont {Johns}},\ }\href
  {\doibase https://doi.org/10.1016/0375-9474(73)90225-X} {\bibfield  {journal}
  {\bibinfo  {journal} {Nuclear Physics A}\ }\textbf {\bibinfo {volume}
  {202}},\ \bibinfo {pages} {289} (\bibinfo {year} {1973})}\BibitemShut
  {NoStop}%
\bibitem [{\citenamefont {Qiu}(2016)}]{qiu2016}%
  \BibitemOpen
  \bibfield  {author} {\bibinfo {author} {\bibfnamefont {Y.}~\bibnamefont
  {Qiu}},\ }\href@noop {} {Ph.D. thesis},\ \bibinfo  {school} {University of
  Massachusetts Lowell} (\bibinfo {year} {2016})\BibitemShut {NoStop}%
\end{thebibliography}
\end{document}